  \documentclass[aps,pre,reprint,groupedaddress]{revtex4-1}

\usepackage{amsmath,amsthm}
\usepackage{amssymb}
\usepackage{hyperref}
\usepackage{graphicx}
\usepackage{breqn} 
\usepackage{bbm}
\newtheorem{theorem}{Theorem}

\newtheorem{lemma}[theorem]{Lemma}

\newtheorem{remark}{Remark}[section]

\usepackage{color}

\newcommand{\df}{\textrm{d}}
\newcommand{\E}[2][n]{\mathbb{E}_{#1} \left[ {#2} \right]}
\newcommand{\EE}{\mathbb{E}_n}
\newcommand{\N}{\mathcal{N}}

\newcommand{\bbT}{\mathbb{T}}
\newcommand{\bbZ}{\mathbb{Z}}
\newcommand{\bbS}{\mathbb{S}}

\DeclareMathOperator{\Tr}{Tr}

\newcommand{\Ito}{It\^o }

\newcommand{\sigi}{ {\boldsymbol \sigma} }
\newcommand{\sigiq}{ \sigma }
\newcommand{ \sig }{ {\vec \sigma} }

\newcommand{\sigpi}{ {\boldsymbol {\tilde \sigma}} }
\newcommand{\sigpiq}{ {\tilde \sigma} }
\newcommand{ \sigp }[1]{ {\vec {\tilde \sigma}^{#1}}}

\newcommand{ \wi }{ {\boldsymbol w} }
\newcommand{ \w }{ {\vec w} }
\newcommand{ \wiq }{ {w} }
\newcommand{ \wq }{ {\boldsymbol w} }

\newcommand{ \nui }{ {\boldsymbol \nu} }
\newcommand{ \ui }{ {\boldsymbol u} }
\newcommand{ \phib }{ \boldsymbol{\zeta} }

\newcommand{ \si }{ {\boldsymbol s} }
\newcommand{ \s }{ {\vec s} }
\newcommand{ \siq }{ {s} }

\newcommand{ \dWi }{ {\mathrm{d} \boldsymbol W} }
\newcommand{ \dWq }{ {\mathrm{d} \boldsymbol W} }
\newcommand{ \dWiq }{ {\mathrm{d} W} }
\newcommand{ \dW }{ {\mathrm{d}\vec W} }
\newcommand{ \Wi }{ { \boldsymbol W} }

\newcommand{ \dt }{ {\mathrm{d} t} }

\begin{document}


\title[Colored Noise in the Landau-Lifshitz System]{Non-local SPDE limits of spatially-correlated-noise driven spin systems derived to sample a canonical distribution}


\author{Yuan Gao}
\email{gaoyuanmath@gmail.com} 
\affiliation{University of North Carolina at Chapel Hill}
 \author{Jeremy L. Marzuola}
 \email{marzuola@math.unc.edu}
\affiliation{University of North Carolina at Chapel Hill}
\author{Katherine A. Newhall}
\email{knewhall@unc.edu}
\affiliation{University of North Carolina at Chapel Hill}
\author{Jonathan Mattingly}
\email{jonm@math.duke.edu}
\affiliation{Duke University}


\date{\today}

\begin{abstract}

We study the macroscopic behavior of a stochastic spin ensemble driven by a discrete Markov jump process motivated by the Metropolis-Hastings algorithm where the proposal is made with spatially correlated (colored) noise, and hence fails to be symmetric.  However, we demonstrate a scenario where the failure of proposal symmetry is a higher order effect.  Hence, from these microscopic dynamics we derive as a limit as the proposal size goes to zero and the number of spins to infinity,  a non-local stochastic version of the harmonic map heat flow (or overdamped Landau-Lipshitz equation).  The equation is both mathematically well-posed and samples the canonical/Gibbs distribution related to the kinetic energy.  The failure of proposal symmetry due to interaction between the confining geometry of the spin system and the colored noise is in contrast to the uncorrelated, white-noise, driven system.  Specifically, the choice of projection of the noise to conserve the magnitude of the spins is crucial to maintaining the proper equilibrium distribution.  Numerical simulations are included to verify convergence properties and demonstrate the dynamics.  
\end{abstract}


\maketitle

\section{Introduction}

In order to accurately describe noise-induced phenomenon in spatially-extended systems, it is important to add fluctuations to continuum models that respect some underlying structure like a Hamiltonian and the sampling of the Gibbs/Boltzmann/canonical distribution.  Guaranteeing this kind of fluctuation-dissipation relation (a.k.a.~detailed balance) is not always obvious, especially in condensed matter physics for which accurate phenomenological models are not always built from first principles.  
One example is the Landau-Lifshitz-Gilbert equation describing a single magnetic spin requiring multiplicative noise, thereby creating an effective electric field, rather than additive noise to ensure sampling of the Gibbs distribution, see \cite{kohn2005magnetic}.  In effect, the noise is projected onto the surface of the sphere representing the configuration space of the constant magnitude spin vector. Another example is the regularization of Stochastic partial differential equations (SPDEs) by correlating the noise in space.  While the 
corresponding dynamics occur at regularity scales that allow for analysis of the evolution to be treated via now well-understood methods for understanding stochastic paths in the PDE setting, see for instance \cite{de1999stochastic,da2014stochastic}, entirely different distributions from their un-correlated noise counterparts may be sampled.  Although white-noise solutions to SPDEs in situations with much less regularity can be understood with the introduction of regularity structures by Hairer in \cite{Hairer:2014hd}, there are still dimensional restrictions, even in the case where the deterministic part is parabolic and hence strongly coercive, see for instance the recent work \cite{bruned2019geometric} on stochastic harmonic map heat flows.
Our goal in this work is to combine the two considerations above related to sampling and regularization, deriving an SPDE model for a spatially-extended magnetic spin system with spatially ``colored'' noise designed to sample an invariant Gibbs measure.


We derive such a continuum model designed to sample an invariant Gibbs measure from a microscopic Metropolis Hastings (MH) algorithm.  The MH algorithm \cite{hastings1970monte,metropolis1953equation} allows the random walk dynamics to be separated from the Hamiltonian structure in the invariant measure:  a simple random-walk proposal, 
$\tilde X_i = X_i^n + \varepsilon w_i^n$ with $i=1\dots N$ indexing space and the $w_i^n$ independent normally distributed random variables,  
 will sample the Gibbs measure
\begin{equation}\label{eq:Gibbs}
\mu(\vec X) = Z^{-1} e^{-\beta H(\vec X)},
\end{equation}
where $Z$ is the partition function and $\beta^{-1}=k_BT$, if an accept probability of
$$
\alpha = 1\wedge e^{-\beta ( H(\tilde{\vec X}) - H(\vec X^n))}
$$
where $a\wedge b = \min(a,b)$ is used for arbitrary bounded Hamiltonian $H$ (i.e.~$\vec X^{n+1} = \tilde{\vec X}$ with probability $\alpha$ and $\vec X^n$ otherwise).  The stochastic differential equation (SDE)
$$
d\vec X = - \nabla H dt + \sqrt{2\beta^{-1}} d\vec W
$$
also samples the invariant measure \eqref{eq:Gibbs}.  Furthermore the MH dynamics converge to the SDE dynamics in the limit as the proposal size $\varepsilon\to 0$.  Thus the limiting MH dynamics can be used to construct SDE models that preserve the invariant measure \eqref{eq:Gibbs} in more complex situations.   For example, if the random walk proposal is changed to  $\tilde{\vec X} = \vec X^n + \varepsilon B \vec w^n$ for constant matrix $B$, then the SDE 
\begin{equation}\begin{aligned}\label{eq:SDEgeneric}
 d\vec X & = - BB^T \nabla H dt + \sqrt{2\beta^{-1}} B  d\vec W
\end{aligned}\end{equation}
also samples the invariant measure \eqref{eq:Gibbs} (but not for every non-constant $B(\vec X)$, c.f.~\cite{PhysRevE.76.011123}).  In  Appendix \ref{App:FP_Addative} we confirm this by direct substitution into the (constant $B$) Fokker-Planck equation
\begin{equation}\begin{aligned}
\label{eq:constB_FP}
\partial_t \rho(x,t) &=  \sum_{i=1}^{3N}  \partial_i \left[ ( BB^T \nabla H )_i \rho(x,t)\right] \\
& + \beta^{-1} \sum_{i,j =1}^{3N}  (BB^T)_{ij} \partial_i \partial_j  \rho(x,t) .
\end{aligned}\end{equation}
Equation \eqref{eq:SDEgeneric}, with symmetric, non-negative definite covariance matrix $BB^T$, has spatially-correlated noise 
 and still samples the Gibbs distribution \eqref{eq:Gibbs}.  A continuum limit of the SDE \eqref{eq:SDEgeneric} exists if the Hamiltonian $H$ and covariance matrix $BB^T$ are appropriately scaled with system size $N$.

In this work, we consider a system of $N$ spins (with periodic boundary conditions), or vectors on $\bbS^m$ for some $m \geq 1$, thereby introducing a confining geometry and investigate how this interacts with spatially-correlated ``colored'' noise, deriving an appropriately regularized Stochastic partial differential equation that still samples an invariant measure of the form \eqref{eq:Gibbs}.   The spatially correlated noise coupled to the geometric constraint will result in a proposal of the form $\tilde{\vec X} = \vec X^n + \varepsilon B (\vec X^n) \vec w^n$, where unfortunately {\it the colored noise proposal is no longer symmetric}.  However, we prove that the MH dynamics can still be approximated by an SDE system similar to that of \eqref{eq:SDEgeneric}, and that for a canonical choice of the matrix $B$ related to the geometry, that the SDE system still samples the correct invariant measure.

For ease of exposition and physical importance, we will restrict ourselves to $m =2$ and work only with spins defined as vectors on $\bbS^2$.  
We build on our recent work \cite{gao2018limiting} which showed that on a general torus in any dimension, $\bbT^d$, the MH dynamics for a system of spatially-uncorrelated ``white'' noise driven spins with confining geometry converged to the dynamics of an SDE system as $\varepsilon$, the proposal size, went to zero.    We also considered the $N\to\infty$ limit of the dynamics while quenching the noise ($\beta =N^\gamma$ for $\gamma$ sufficiently large) to arrive at the harmonic map heat flow equation 
\begin{equation}\label{eq:whitePDE}
\partial_t \sigma(x,t) = - \sigma \times ( \sigma \times \Delta \sigma). 
\end{equation}  
This could also be referred to as the overdamped Landau-Lifshitz-Gilbert (LLG) equation.  
Quenching the noise was essential in the derivation due to the known convergence issues with stochastic partial differential equations (SPDE) in spatial dimensions greater than one (c.f.~\cite{ryser2012well}).   The convergence from the SDE model to a PDE model also relied on the regularity of the harmonic map heat flow equation, which can fail for $\bbT^d \to \bbS^m$ in finite time for dimensions $d > 2$ due to bubbling singularities, see \cite{struwe1988evolution,guo1993landau}.

To derive a regularized SPDE limit ($\beta$ constant with $N\to\infty$), we begin with a random walk for the MH algorithm that projects now spatially-correlated Gaussian noise onto the tangent plane of the underlying geometry.  After taking the proposal size $\varepsilon\to0$ arriving at a system of SDEs, we find that
unlike the white noise case, the choice of $\sigma \times$ as the projection is crucial for sampling the desired distribution \eqref{eq:Gibbs}.  Therefore, the regularized non-local SPDE that samples the Gibbs distribution is 
\begin{align}
\label{eq:spde}
\partial_t \sigma(x,t) & =  -\sigma(x,t) \times \int_{\bbT^d} C(x-y)  (\sigma \times \Delta \sigma) (y,t)  dy   \nonumber \\
& + \sqrt{2\beta^{-1}}  \sigma(x,t) \times  \eta^C(x,t),
\end{align}
where $C$ is a non-local operator to be described below in a variety of cases that encodes the covariance structure of the colored noise, $\eta^C(x,t)$, is colored-in-space white-in-time noise (i.e. $\mathbb{E}[ \eta^C(x,t)\eta^C(y,s)] = C(x-y)\delta(t-s)$), interpreted in the Stratonovich sense.

\subsection{Prior Work}

Having an appropriately regularized stochastic limit is important to studying thermal effect in ferromagnets such as magnetization reversal~\cite{Wernsdorfer:1997eb,Coffey:2012ez}.  Existing field models continue to use spatially-uncorrelated white noise in the stochastic LLG equation so as to maintain the equilibrium distribution, proposing for example weak formulations of the solutions and numerical finite element schemes (c.f.~Ch.~2 of \cite{banas2014stochastic}).
Equation \eqref{eq:spde} is in contrast to regularizing the LLG equation by changing the energy functional to include a term to control the modulus of continuity~\cite{Chugreeva:2018iz}.
It also compliments other works that derive equations to preserve the equilibrium distribution, such as in the case of inhomogeneous magnitude of magnetic spins~\cite{Nishino:2015gv}, for temporally-colored noise but for finitely many spins~\cite{Atxitia:2009io}, and for the stochastic Landau-Lifshitz-Bloch equation~\cite{Evans:2012ex}.   More generally, physical models with confining geometries are natural generalizations of the SPDE limits derived using colored noise for unconstrained random walks in \cite{MPS} and more recently in \cite{Kuntz:2019fh}.  See also \cite{PhysRevE.60.6343} that focuses on quantum field theories, but also discusses 
 the effective action of a generic SPDE system through the tools of fluctuation-dissipation and invariant measures, with examples including reaction-diffusion-decay systems, KPZ (noisy Burgers), and purely dissipative SPDEs.

Since our approach starts from the MH algorithm, it is worth pointing out that
the MH algorithm itself  is widely used in particle statistics and sampling algorithms, see for instance \cite{newman1999monte,binder1993monte,landau2014guide,batrouni2004metastable,maccari2016numerical}.  It also arrises when adopting the Bayesian approach to inverse problems and signal processing~\cite{stuart2010inverse,hairer2011signal}.  This has lead to the study of optimal scalings for the unconstrained random walk MH algorithm and diffusion limits  for certain forms of probability distributions~\cite{roberts1997weak,breyer2000metropolis,MPS,jourdain2014optimal,jourdain2015optimal,Kuntz:2019fh}.  Specifically, for product measures in \cite{roberts1997weak} and the Gibbs distribution of a lattice model in \cite{breyer2000metropolis}, the weak convergence to Langevin diffusions has been shown by comparing generator functions. The pioneering work \cite{MPS}, based in part upon earlier works on sampling \cite{hairer2005analysis,hairer2007analysis}, extended this type of result to non-product form measures  and demonstrated the weak convergence to a SPDE.  Subsequent works \cite{jourdain2014optimal,jourdain2015optimal,Kuntz:2019fh} consider scaling limits of systems started away from their equilibrium distributions.

  Building on our previous work \cite{gao2018limiting} that studied the limiting dynamics of a geometric MH process with white noise in the proposal,  
we fill a missing gap in the above results showing strong convergence of trajectories started far from equilibrium to a non-local SPDE in a geometric setting,  
 with the underlying dynamics of the process designed to sample an (non product form) invariant measure using colored noise with a given covariance structure. 
Similar to  \cite{MPS}, we derive a  drift term that implicitly is driven by a non-local diffusion operator.  In the context of random walks, this is related to fractional diffusion operators, but we are interested to see the effects of colored noise on the geometric evolution.

\subsection{Outline of Results}

The remainder of the paper is as follows.  In section~\ref{sec:notation} we layout the vector notation we adapt for the paper.  In section~\ref{sec:white_noise} we review our results from \cite{gao2018limiting} pointing out a few interesting facts that will be in contrast to the colored noise case.   We extend these results to the case of colored noise in section \ref{sec:colored_noise}, outlining the derivation of the limiting SDE system from the MH dynamics in section \ref{sec:MHtoSDE} (details of the proof in Appendix \ref{app:SDEderivation}), discussing the correct projection of the noise onto the tangent plane of the underlying geometry to ensure the SDE system samples the desired distribution \eqref{eq:Gibbs} in section \ref{sec:cross}, proving the invariant measure of the MH dynamics converges to this same invariant measure in section \ref{sec:IM},  and discuss the Fourier representation of the non-local SPDE \eqref{eq:spde} in section \ref{sec:PSDE} with an outline the well-posedness in Appendix \ref{A:LWP} when the noise is trace class. We support our trajectory-wise convergence results with direct numerical simulations in section \ref{sec:num} as well as illuminate the differences between the choice of two different projections of the noise onto the tangent plane of the underlying geometry.   We give concluding remarks in section \ref{sec:conclusions}.

\section{Notation\label{sec:notation}}

We present our results for the case of one periodic spatial dimension, $\bbT$, and spins that live on $\mathbb{S}^2$, although this can be extended to other dimensions for both the periodic domain and the spherical target.  It becomes convenient  to adopt different notation in different contexts, which we summarize here.  The torus with unit length is subdivided with $x_i = (i-1)/N$ for $i=1\dots N$ with a spin located at each $x_i$.  We take $\sigi_i^n$ for $i=1\dots N$ as the collection of the $N$ spins of the MH dynamics at time-step $n$, each a 3-dimensional vector, with components 
\begin{equation}
\sigi_i^n = \left< \sigiq_{i,x}^n, \sigiq_{i,y}^n, \sigiq_{i,z}^n \right>
\end{equation}
 satisfying $( \sigiq_{i,x}^n)^2 + (\sigiq_{i,y}^n)^2+ (\sigiq_{i,z}^n)^2 = 1$ for each $i=1\dots N$ and each integer $n\ge 0$.  The $3N$-dimensional vector 
\begin{equation}
\sig^n = \left< \sigiq_{1,x}^n \dots \sigiq_{N,x}^n\;  \sigiq_{1,y}^n \dots \sigiq_{N,y}^n\;  \sigiq_{1,z}^n \dots \sigiq_{N,z}^n \right>
\end{equation}
contains all the components of all the spins.  We similarly define: $\sigp{n}$, $\sigpi_i^n$ and $\sigpiq_{i,q}^n$ $q\in\{x,y,z\}$ for the MH proposal at time-step $n$; $\w^n$, $\wi_i^n\in\mathbb{R}^3$, and $\wiq_{i,q}^n$ for the independent standard Gaussian random variables used to generate the proposal at time-step $n$; $\s(t)$, $\si_i(t)$, $\siq_{i,q}(t)$ for the solution to the limiting SDE system at time $t$.  Since the noise will be correlated in each component, it will also be useful to represent it as
$$
\w^n = \left< \wq_x^n \;  \wq_y^n \;  \wq_z^n \right>
$$
with each $N\times 1$ vector
$$
\wq_q^n = \left< \wiq_{1,q}\dots \wiq_{N,q} \right> \textrm{ for } q\in\{x,y,z\}.
$$

\section{White Noise} \label{sec:white_noise}

Here we present an overview of our previous work, \cite{gao2018limiting}, pointing out a few interesting facts that will be in contrast to the colored noise case.  We remind the reader that though we limit our discussion here to the cases $d=1$, $m=2$ for ease of exposition, all results here extend to $d\geq1$, $m \geq 1$ with small modifications.

To arrive at an appropriate continuum limit, we begin with the standard Metropolis Hastings algorithm using independent  Gaussian (``white'') noise to propose a new state.  The proposed new configuration of the $N$ spins $\sigpi_i^n$, $i=1,2,\dots N$ requires picking a random direction in the tangent plane, moving along that direction, and projecting back onto the sphere, 
\begin{equation}\label{eq:proposal}
\sigpi^{n}_i = \frac{  \sigi_i^n + \epsilon \nui_i^W }{ \| \sigi_i^n + \epsilon \nui_i^W \| }
\end{equation}
with $\nui_i^W = \textrm{P}^\perp_{\sigi_i^n} (\wi_i^n)$ is a projection of the three-dimensional normal random vector $\wi_i^n$ into the tangent plane of $\sigi_i^n$,  $\textrm{P}^\perp_x( y) = y-(x\cdot y)x $ or in matrix form $(I - xx^T)y$.  Defining Hamiltonian 
\begin{equation} \label{eq:H}
 H(\sig) = \tfrac{1}{N} \sum_{i=1}^N  \tfrac{N^2}{2}  \| \sigi_{i+1} - \sigi_i \|^2
\end{equation}
with $\sigi_{N+1} = \sigi_1$ for periodic boundary conditions,
the accept probability 
\begin{equation}\label{eq:alpha}
\alpha = 1\wedge e^{-\beta ( H(\sigp{n}) - H(\sig^n))}
\end{equation}
ensures sampling of the Gibbs distribution \eqref{eq:Gibbs}, where $\sigp{n}$ and $\sig^n$ are the $3N$-vectors of the proposal components and current spin components, respectively.  Symmetry in the proposal is crucial for \eqref{eq:alpha} to be the correct accept probability to sample the Gibbs distribution.  We discuss this in more detail, pointing out that symmetry is lacking when $\nui_i^W$ in the proposal is replaced with its correlated noise version next in Sec.~\ref{sec:colored_noise}.

By taking the lowest order term in $\varepsilon$ of the mean and added noise, the MH step is approximately equivalent to the Euler-like step
$$
 \sigi_i^{n+1} - \sigi_i^n  \approx -\tfrac12 \beta \varepsilon^2 \textrm{P}^\perp_{\sigi_i^n}\left( \frac{\partial H}{\partial \sigi_{i}^n}\right) - \varepsilon^2 \sigi_i^n + \varepsilon \textrm{P}^\perp_{\sigi_i^n} (\wi_i^n).
$$
 Our previous work showed the trajectory-wise convergence as $\epsilon\to 0$ of the MH dynamics to the corresponding It\^o SDE
\begin{equation}\label{eq:whiteSDE}
\mathrm{d}\si_i = \left[ \textrm{P}^\perp_{\si_i}(\Delta_N\si_i) - \tfrac{2N}{\beta} \si_i \right] \dt + \sqrt{2\beta^{-1}N} \textrm P^\perp_{\si_i}( \dWi_i )
\end{equation}
under the time rescaling $\delta t = \beta \varepsilon^2/2N$ where $\Wi_i$ are 3-dimensional Brownian motions and 
\[
\Delta_N \sigi_i = N^2( \sigi_{i+1}-2\sigi_i + \sigi_{i-1})
\] is the discretized Laplace operator.

In the case of white noise, we point out that other projection operators could be used in place of $\textrm P^\perp_x(y)$ above.  The only requirement in the MH algorithm is that white noise is projected onto the tangent plane of $\sigi_i^n$.   Two other natural choices would be $\sigi_i^n \times \wi_i^n$ and $- \sigi_i^n \times (\sigi_i^n \times \wi_i^n)$, the later being equivalent to $\textrm P^\perp_{\sigi_i^n} (\wi_i^n)$ defined above; both produce white noise in the tangent plane.  We will observe in Sec.~\ref{sec:cross} that this freedom is strongly related to the white noise setting and that care must be taken when moving to the colored noise case.

To see the equivalence of the two natural projection choices of the cross and cross-cross product in the white-noise case, we show that the limiting SDE systems for the MH dynamics produce the exact same Fokker-Planck equation in either case, so using either is justified.  Define
the $3N\times 1$ vector of independent noises as
\begin{equation}\label{eq:dWvec}
\dW = \left< \dWq_{x}\;  \dWq_{y} \; \dWq_{z} \right>
\end{equation}
with each $N\times 1$ vector
$$
\dWq_q = \left< \dWiq_{1,q}\dots \dWiq_{N,q} \right> \; \textrm{for } q\in\{x,y,z\}
$$
so that the $3N$  (\Ito) equations analogous to \eqref{eq:whiteSDE} are
\begin{equation}\begin{aligned}\label{eq:ItoWhite}
\mathrm{d}\s =  PP^T \Delta_N\s \;\dt - \frac{2N}{\beta} \s \;\dt + \sqrt{2\beta^{-1}N} P \dW.
\end{aligned}\end{equation}
  We consider two choices for the block-defined projection matrix $P$ next.  Note that both these projection matrices contributes the same factor $-2\s$ to the \Ito correction term, $-2N\beta^{-1}\s$, in the above SDE.
For the single-spin projection $\sigi_i \times \dWi_i$,  
the block-defined projection matrix is  
\begin{equation}\label{eq:Pcross}
P_1 = \begin{pmatrix}
        0 & -Z & Y \\ 
        Z & 0 & -X \\
        -Y & X & 0
       \end{pmatrix}
\end{equation}
and the block-defined projection matrix for $-\sigi_i \times (\sigi_i \times \dWi_i)$  is
\begin{equation}\label{eq:Pcrosscross}
 P_2 = \begin{pmatrix}
        I-X^2 & -XY & -XZ \\
        -XY & I-Y^2 & -YZ \\
        -XZ & -YZ & I-Z^2
       \end{pmatrix}
\end{equation}
where each $N \times N$ block matrix $X,Y$ or $Z$ are the diagonal matrices 
\[ Q = 
  \begin{pmatrix}
    \sigiq_{1,q} & 0 & \dots & 0 \\
    0 & \sigiq_{2,q} & \dots & 0 \\
    \vdots & \vdots & \ddots & \vdots \\
    0 & 0 & \dots & \sigiq_{N,q}
  \end{pmatrix}
\]
for $Q\in\{X,Y,Z\}$ with corresponding $q\in\{x,y,z\}$.  The Fokker-Planck equation for \eqref{eq:ItoWhite} is
\begin{equation}\begin{aligned}
\partial_t \rho(\s,t) &=  \sum_{i=1}^{3N}  \partial_i \left[  (PP^T \Delta_N \s )_i \rho(\s,t)\right] \\
 & + \frac{2N}{\beta}\sum_{i=1}^{3N}  \partial_i \left[ \si_i \rho(\s,t) \right]\\
& + \frac{N}{\beta} \sum_{i,j =1}^{3N} \partial_i \partial_j \left[ (PP^T)_{ij} \rho(\s,t) \right].
\end{aligned}\end{equation}
Notice that this equation only depends on  $PP^T$
 which is identical for both 
 $P_1$ and $P_2$,
\begin{equation}\label{eq:equivalent}
 P_1 P_1^T = P_2 P_2^T = P_2,
\end{equation}
after using that $\sigiq_{i,x}^2 + \sigiq_{i,y}^2 + \sigiq_{i,z}^2 =1$ for each $i$.  Thus, both projections produce statistically equivalent trajectories in the white noise setting, and direct substitution can verify that \eqref{eq:Gibbs} is an invariant measure for both (see Appendix \ref{App:FP_Mult}).  The key point when taking colored noise instead of white noise that we will see in Sec.~\ref{sec:cross} is that {\em the covariance matrix for the noise and the projection matrix do not commute} and therefore $PP^T$ does not appear isolated in the colored noise Fokker-Planck equation.  The two projection matrices $P_1$ and $P_2$ produce statistically different ensembles.

We also point out that the \Ito correction term in \eqref{eq:ItoWhite}, $-2N\beta^{-1}\s$, is completely independent of choice of projection, the Stratonovich form of \eqref{eq:ItoWhite} being  
\begin{equation}\begin{aligned}\label{eq:whiteSDEstrat}
\mathrm{d} \s =  PP^T \Delta_N\s \;\dt + \sqrt{2\beta^{-1}N} P \circ \dW.
\end{aligned}\end{equation}
This fact remains true in the case of colored noise, that the \Ito correction term depends only on the covariance matrix of the noise but not the choice of projection (see Sec.~\ref{sec:colored_noise} with details in Appendix \ref{append:ito_correction}).

Our previous work also considered the continuum limit of the SDE system \eqref{eq:whiteSDE}.  Defining a lattice spacing $\delta x = N^{-1}$ and taking a scaling of $\beta = N^{\gamma}$ for $\gamma$ sufficiently large to quench the noise (numerical simulations verified convergence for $\beta \sim N^{3/2}$), we showed convergence to the (local, deterministic) PDE
\eqref{eq:whitePDE} under some regularity assumptions of the solution to the harmonic map heat flow equation.  This convergence holds regardless of the number of spatial dimensions considered, provided we assume regularity of the solution to the corresponding harmonic map heat flow with domain $\mathbb{T}^d, d>2$.  As mentioned in the Introduction, the regularity of the solution for $d > 2$ is a delicate issue when considering the fixed $\beta$ continuum limit to an SPDE and one may not be guaranteed convergence in the case of white noise.  

\section{Colored Noise} \label{sec:colored_noise}

Using spatially-correlated noise in the proposal of the MH algorithm to lead to regularized SPDEs in the continuum limit intuitively accounts for the fact that at smaller atomic scales, the true physical system cannot be further subdivided into infinity small units with independent fluctuations.  A natural way to introduce correlations in the noise that decay with distance is to ``color'' the noise, requiring the power in the Fourier representation to decay with frequency.  In the discrete setting, 
to form various covariance matrices satisfying our periodic boundary conditions, we use a periodic Fourier basis with power in each frequency mode that decays with rate
 $\kappa$.  We again remind the reader that for ease of exposition we have set $d=1$ in this section, but extending to higher dimensions is just a matter of using higher dimensional discrete Fourier transform machinery.  However, in subsection \ref{sec:PSDE} below about SPDE limits, we will state the limiting equations for general dimension $d$.

Specifically we decompose an $N\times N$ covariance matrix 
 \begin{subequations}\label{Cbar}
 \begin{equation}
 \bar{C}_N = \phi \bar{D}^2 \phi^T
 \end{equation}
 with diagonal matrix $\bar{D}_{jj} = \lambda_j = d_j^{-\kappa}$ with frequencies $d_j$ defined as
\begin{equation}\label{Dbar}
d_j = \left\{ \begin{array}{ll} 
1 & j=1 \\
2\pi (j-1) &  2\le j \le \tfrac{N}{2} + 1 \\
2\pi(j-\tfrac{N}{2} - 1) &   \tfrac{N}{2} + 2 \le j \le N
\end{array}\right.
\end{equation}
and the matrix of Fourier eigenvectors given by 
\begin{equation}\label{Phi}
\phi_{ij} =  \left\{ \begin{array}{ll}
1 & j=1 \\
\sqrt{2} \cos(\frac{1}{N} d_j (i-1) ) & 2\le j \le \tfrac{N}{2} \\
 \cos(\frac{1}{N} d_j (i-1) ) & j = \tfrac{N}{2} + 1\\
\sqrt{2} \sin( \frac{1}{N} d_j (i-1) )  & \tfrac{N}{2}+2 \le j \le N   
\end{array}\right.     
\end{equation}
\end{subequations}
with $ \sum_i \phi_{ij}^2 = N $ for each $j$.  With this scaling, the eigenvectors $\phi_{ij}$ converge to the discrete set of Fourier functions $1$, $\sqrt{2}\cos(2\pi x), \sqrt{2}\cos(4\pi x), \dots $ and $\sqrt{2}\sin(2\pi x), \sqrt{2}\sin(4\pi x)\dots $ as $N\to\infty$, forming an orthonormal set, with inner product of two functions defined as $\int_0^1 f(x) g(x) dx $. 
Also due to this scaling $\Tr(\bar{C}_N) = N \sum_{j=1}^N \lambda_j^2$.  Note that taking $\kappa=0$ creates equal power in all modes, reducing $\bar{C}_N$ to a diagonal matrix with $N$ on the diagonal representing uncorrelated ``white'' noise.  Increasing $\kappa$ increases the length scale of the covariance, broadening $\bar{C}_N$ which is peaked along the diagonal.   

For use in the MH algorithm, at time step $n$, we form three vectors $\bar{C}_N^{1/2} \wq_q^n$ for $q\in\{x,y,z\}$ with the $\wq_q^n$ a set of vectors of independent uncorrelated standard Gaussian random variables.  The vectors $\bar{C}_N^{1/2} \wq_q^n$ are independent for different $q$ but spatially-correlated with covariance matrices given by  $\bar{C}_N$.  This correlated noise is projected into the tangent plane of the corresponding spin, defining
$$
\nui_i^n = P^\perp_{\sigi_i^n} ( (\bar{C}_N^{1/2}\wq^n_x)_i, (\bar{C}_N^{1/2}\wq^n_y)_i, (\bar{C}_N^{1/2}\wq^n_z)_i ).
$$ 
The analogous proposal to \eqref{eq:proposal} is
\begin{equation}\label{eq:proposalcolor} 
\sigpi_i^n = \frac{  \sigi_i^n + \varepsilon \nui_i^n }{ \| \sigi_i^n + \varepsilon \nui_i^n \| }.
\end{equation}

The first thing to note is that using $\nui_i^n$ in place of $\nui^W_i$ in the proposal  creates a non-symmetric proposal and therefore using the accept probability \eqref{eq:alpha}  no longer guarantees sampling of the Gibbs distribution \eqref{eq:Gibbs}.  In particular, since our noise is now spatially correlated but our projections are completely local, the probability of undoing a particular rotation is not equal to the probability of doing it.  
In the white noise case, the tangent vector $\tilde{\nui}_i^W$ to get $\sigi_i^n$ back from the proposal $\sigpi_i^n$ is unique and has the same magnitude as $\nui_i^W$. Then $\mathbb{P} (\sigi_i^n | \sigpi_i^n ) = \mathbb{P} (\sigpi_i^n | \sigi_i^n )$ and since the tangent vectors $\nui_i^W$ are independent for different spins $i$, the entire proposal in the white noise case is symmetric,
	\[
		\mathbb{P} (\sig^n | \sigp{n} ) = \mathbb{P} (\sigp{n} | \sig^n ).
	\]
In the colored noise case, the tangent vectors are correlated and the above symmetry requirement is no longer true.  However, as the sphere is locally close to flat, intuitively the projected tangent vectors from $\sigi_i^n$ and back from the proposal $\sigpi_i^n$ should be almost symmetric, though we observe that it depends upon the projection chosen as to how this asymmetric proposal manifests in the limit of $\epsilon\to 0$.  For the cross-product projection corresponding to $P_1$, the non-symmetric terms appear in higher-orders of $\varepsilon$ and we conjecture they vanish taking similar limits of the (wrongly defined) MH algorithm as we did previously.   We revisit this conjecture in Sec.~\ref{sec:IM}.

We discuss this limit of $\varepsilon\to0$, arriving at the (It\^o) SDE 
\begin{equation} \label{eq:colorSDE_ito}
\begin{aligned}
  \mathrm{d} \s =&    P_1 \frac{C_{N}}{N}  P_1^T \Delta_N\s \;\dt  - 2\beta^{-1} \frac{\textrm{Tr}( \bar{C}_N )}{N} \s \;\dt  \\
 &  + \sqrt{2\beta^{-1}} P_1 C_{N}^{1/2} \dW
\end{aligned}
\end{equation}
next in Sec.~\ref{sec:MHtoSDE} with details appearing in Appendix \ref{app:SDEderivation}.   Then in Sec.~\ref{sec:cross} we discuss why the $P_1$ projection matrix, corresponding to $\sigma \times $ has been selected.  In Appendix  \ref{App:FP_Mult} we verify that the Gibbs distribution is the invariant measure of \eqref{eq:colorSDE_ito} by considering the Fokker-Planck equation for the equivalent Stratonovich SDE
 \begin{equation}\label{eq:colorSDE}
 \begin{aligned}
 \mathrm{ d} \s =&  P_1 \frac{C_{N}}{N} P_1^T \Delta_N\s \;\dt   + \sqrt{2\beta^{-1}} P_1 C_{N}^{1/2} \circ \dW. 
\end{aligned}\end{equation}
Notice that for the case of uncorrelated noise, $\kappa=0$, the matrix $C_N$ reduces to a diagonal matrix with $N$ on the diagonal.  The SDE \eqref{eq:colorSDE_ito} therefore reduces to the white noise SDE \eqref{eq:whiteSDE} as $\frac1N C_N$ reduces to the identity matrix, $\frac1N \Tr(\bar{C}_n) = N$ and $C^{1/2}_N \dW = \sqrt{N} \dW$.

\subsection{Limiting Dynamics of Metropolis Hastings\label{sec:MHtoSDE}}

The idea behind the convergence is to equate one Metropolis Hastings step to one Euler-Maruyama numerical integration step of the It\^o SDE \eqref{eq:colorSDE_ito}.  Following  \cite{gao2018limiting,MPS} we consider the leading order in proposal size $\varepsilon$ terms for the drift and diffusion of one MH step.  At various points we drop higher order terms that are random variables, which are capable of taking on arbitrarily large values, but with small probability.  To ensure a true asymptotic convergence, we bound the average pathwise error between MH and SDE trajectories themselves, not the probability distribution governed by a master equation, thus we have a strong, trajectory-wise, convergence result.  In what follows, we heuristically explain obtaining the leading order terms for the drift and the diffusion.  Expectations, $\E{\cdot}$, are conditioned on knowing the current MH spin configuration, $\sig^n$.   The details of properly bounding the error between the piece-wise interpolated MH trajectory and the SDE trajectory are left to Appendix \ref{app:SDEderivation}.

The drift term of the SDE comes from the expectation of one MH step,
\begin{equation}\label{eq:expDrift}\begin{aligned}
& \E{ \sig^{n+1} - \sig^n } =\\
&\hspace{1cm} \E { (\sigp{n} - \sig^n) \left(1 \wedge e^{-\beta \big( H(\sigp{n}) - H(\sig^n) \big) }\right) },
\end{aligned}\end{equation}
where the elements of the proposal $\sigp{n}$ are each given by \eqref{eq:proposalcolor}.  Expanding this proposal for small $\varepsilon$, we obtain
\begin{equation}\label{eq:expansion}
\sigpi_i^n - \sigi_i^n \approx \varepsilon \nui_i^n - \frac 12 \varepsilon^2 \|\nui_i^n\|^2 \sigi_i^n.
\end{equation}
We evaluate the expectation in \eqref{eq:expDrift} for the first term on the right-hand-side of \eqref{eq:expansion} first, then the second term.

For the expectation over the first term in the expansion \eqref{eq:expansion}, we have
$$
\sigp{n} -\sig^n \approx \varepsilon P C_N^{1/2} \w^n
$$ 
and proceed to compute
$$
\E{ \varepsilon P C_N^{1/2} \w^n  \left(1 \wedge e^{-\beta \big( H(\sigp{n}) - H(\sig^n) \big) }\right) }
$$
using the first order expansion of  $H(\sigp{n}) - H(\sig^n)$ which is
\begin{equation}\label{deltaH}
 \delta H \approx  \varepsilon (\nabla H)^T P C_N^{1/2} \w^n.
\end{equation}
The first order term in the expansion of $\left(1 \wedge e^{-\beta  \delta H }\right)$ is one, resulting in the expectation of $\w^n$ which is zero.  The next order term comes from using the  Lemma 2.4 in \cite{MPS}, which we state here for convenience.
\begin{lemma}[\cite{MPS}] 
\label{lm:mps}
For $z \sim \mathcal{N}(0,1)$,
\begin{equation*}
\E[]{z \left( 1 \wedge e^{az+b} \right) } = a e^{\frac{a^2}{2} + b} \Phi \left( - \frac{b}{|a|} - |a|  \right)
\end{equation*}
for any real constants $a,b$, and $\Phi(\cdot)$ is the CDF for the standard normal random variable.
\end{lemma}
We apply this lemma on the expectation for a single component of $\w^n$ and corresponding coefficient of $\delta H$ and then take the expectation over the remaining components of $\w^n$ with further approximations detailed in Appendix \ref{App:Drift}.  The result is the same as taking $1\wedge e^{-\beta\delta H}\approx 1 - \beta\delta H$ when $\delta H <0$ and 1 otherwise while also assuming $\delta H$ is mean zero so that each case happens with probability 1/2.  Thus,
$$
  \E{  \w^n  \left(1 \wedge e^{-\beta \big( H(\sigp{n}) - H(\sig^n) \big) }\right) } 
 \approx  \E{  \w^n  \left( \frac 12 \delta H \right) } 
$$
and the only expectation that remains is $\E{ \w^n( \w^n)^T } = I$, the identity matrix.  Therefore, 
\begin{equation}\label{w_lemma2}
\E{ \w^n  \left(1 \wedge e^{-\beta \delta H} \right) } \approx - \varepsilon \frac{\beta}{2} ((\nabla H)^T P C_N^{1/2} )^T
\end{equation}
and 
\begin{equation}\label{eq:2}
\E{ \varepsilon P C_N^{1/2} \w^n \left(1 \wedge e^{-\beta \delta H} \right) } \approx- \varepsilon^2 \frac{\beta}{2} PC_NP^T \nabla H.
\end{equation}

Returning to \eqref{eq:expDrift}, we consider the second term in the expansion \eqref{eq:expansion}, and compute  
$$
 \E{\frac 12 \varepsilon^2 \|\nui_i^n\|^2 \sigi_i^n  \left(1 \wedge e^{-\beta \delta H }\right)}.
$$
Here, it is convenient to take $\nui_i^n=P^\perp_{\sigi_i^n} \ui_i$ and write  the three components of $\ui_i$ in terms of the decomposition of the matrix $\bar{C}_N$ defined in \eqref{Cbar} as 
\begin{equation}\label{eq:u}
u_{i,q} = \sum_{j=1}^N \lambda_j \phi_{ji} \wiq_{j,q}^n \qquad \textrm{for } q\in \{x,y,z\}.
\end{equation}
 Unlike above, the first term in the expansion of $\left(1 \wedge e^{-\beta  \delta H }\right)$ gives non-zero expectation, which is
$$
 \E{\frac 12 \varepsilon^2 \|\nui_i^n\|^2 \sigi_i^n } = 
  \varepsilon^2  \sigi_i^n \sum_{j=1}^N  \lambda_j^2 \phi_{ji}^2 .
$$
We further notice that $\sum_{j=1}^N \lambda_j^2 \phi_{ji}^2$ is equivalent to $\frac{1}{N} \textrm{Tr}( \bar{C}_N)$ for each $i$ as a result of the chosen Fourier basis to represent $\bar{C}_N$.  Therefore,
\begin{equation}\begin{aligned}
\label{ito}
 \E{-\frac 12 \varepsilon^2 \|\nui_i^n\|^2 \sigi_i^n } = -\varepsilon^2  \frac{1}{N}\textrm{Tr}( \bar{C}_N) \sigi_i^n.
\end{aligned}\end{equation}
In vector form, combining the above with \eqref{eq:2}, we have that \eqref{eq:expDrift} to leading order in $\varepsilon$ is 
\begin{equation}\label{eq:4}
 \E{\sig^{n+1} - \sig^n} \approx  -\varepsilon^2 \frac{\beta}{2} PC_N P^T \nabla H -  \varepsilon^2 \frac 1N\textrm{Tr}( \bar{C}_N) \sig^n.
\end{equation}

The diffusion part of the SDE is the leading-order in $\varepsilon$ term of the mean-zero noise, 
\begin{equation}\label{eq:diff_approx}
\sig^{n+1} - \sig^n - \E{ \sig^{n+1} - \sig^n} \approx \varepsilon PC_N^{1/2} \w^n.
\end{equation}
Recall from above, that the expectation of $PC_N^{1/2} \w^n$ was zero; this is the leading order noise term.
Combining with the drift, we have that one step of the MH algorithm to leading order is
\begin{equation}\label{eq:SDEstep}
\begin{aligned}
 \sig^{n+1} - \sig^n \approx &-\varepsilon^2 \frac{\beta}{2} P C_N P^T \nabla H \\
 & - \varepsilon^2 \frac{1}{N}\textrm{Tr}( \bar{C}_N )\sig^n
 + \varepsilon PC_N^{1/2} \w^n.
\end{aligned}\end{equation}
Defining a rescaling of time as $\delta t = \varepsilon^2 \beta/ 2 $ the above is 
\begin{equation}\label{eq:SDEstepEM}
\begin{aligned}
 \sig^{n+1} &- \sig^n \approx -  \delta t P \frac{1}{N} C_N P^T \Delta_N \sig^n  \\
 & - \delta t \frac{2}{N \beta}  \textrm{Tr}( \bar{C}_N )\sig^n
 + \sqrt{\frac{2\delta t}{\beta}}  PC_N^{1/2} \w^n,
\end{aligned}\end{equation}
where we have used that for Hamiltonian \eqref{eq:H}, 
\begin{equation}\label{gradHdef}
\nabla H = \tfrac1N \Delta_N \sig^n .
\end{equation}
  Equation \eqref{eq:SDEstepEM} is one step of the 
the Euler-Maruyama method for the Stratonovich SDE \eqref{eq:colorSDE}.

The trajectory-wise convergence of the MH dynamics to the solution of \eqref{eq:colorSDE} is summarized in the following statement and proved in Appendix \ref{App:Existence}.
\begin{theorem} \label{thm:MHtoSDE}
  Define the piecewise constant interpolation of the MH dynamics as $\sig(t)$,
  \begin{equation}
\sig(t) = \sig^n \quad n \delta t \le t < (n+1) \delta t,
  \end{equation}
  where $\delta t = \frac{\beta \varepsilon^2}{2}$ is the timestep size of the MH dynamics, and $\s(t)$ is the solution to the 
   SDE system \eqref{eq:colorSDE} with initial conditions $\s(0) = \sig(0)$ and $\| \sigi_i(0) \| =1,1 \le i \le N$.
  If the proposal noise in the MH step is generated by the same $3N$ Weiner processes in \eqref{eq:colorSDE} as
  $$
   \varepsilon \wi_i^n = \sqrt{2 \beta^{-1}} \left[ \Wi_i((n+1)\delta t) -\Wi_i(n \delta t) \right], 
   $$ 
  for $i=1\dots N$, then we have the following strong convergence result:
  \begin{equation}\label{eq:36}
  \E[]{ \sup_{0 \le \tau \le T} \| \s(\tau) - \sig(\tau)\|^2 } \le c_1 \sqrt{\delta t} \exp(c_2 T)
  \end{equation}
  for any $T \in (0,\infty)$, where $c_1$ and $c_2$ are functions of $N,\beta,T,\Tr(C_N)$ and independent of the choice of $\delta t$.
\end{theorem}

This convergence result holds regardless of the projection matrix $P$.  What remains to be determined is if the SDE \eqref{eq:colorSDE} has the Gibbs distribution \eqref{eq:Gibbs} as its invariant measure.

\subsection{Choosing a Projection\label{sec:cross}}

Having established convergence of the MH dynamics, we are left to show that the system of SDEs \eqref{eq:colorSDE} has the Gibbs distribution \eqref{eq:Gibbs} as its invariant measure.  This SDE is in the form of \eqref{eq:SDEgeneric} with the matrix $B=PC_N^{1/2}$ being non-constant.  For generic non-constant $B$ in \eqref{eq:SDEgeneric}, the Gibbs distribution \eqref{eq:Gibbs} is no longer an invariant measure, however in a few special cases it is.  For example, in the case of the white noise SDE \eqref{eq:whiteSDE} using either $\sigi_i \times \dWi_i$ or $- \sigi_i \times (\sigi_i \times \dWi_i)$, so that $B=P_1$ or $B=P_2$, it is, as we show by direct computation in Appendix \ref{App:FP_Mult}.  However, when considering colored noise, only the projection of the form $\sigi_i \times (\cdot)$ corresponding to $B=P_1 C_N^{1/2}$, and not $B=P_2 C_N^{1/2}$, has \eqref{eq:Gibbs} as an invariant measure, as we show by direct computation in Appendix \ref{App:FP_Mult}.  Unlike the white noise case, since the colored noise matrix and projection matrix do not commute, $P_1 C_N P_1^T \ne P_2 C_N P_2^T$, and the two projections of the noise into the tangent plane produce statistically different trajectories.  We explore this idea further numerically in Sec.~\ref{sec:num}, showing that the cross-cross projection samples something further and further from the Gibbs distribution as the noise becomes more correlated.

 \subsection{Convergence of the Invariant Measure\label{sec:IM}}

In this section, we justify a statement said earlier in Sec.~\ref{sec:colored_noise}, that the non-symmetric terms in the MH proposal \eqref{eq:proposalcolor} appear in higher-orders of the proposal size $\varepsilon$. In particular, we show that the invariant measure of the MH dynamics with colored noise in the proposal and cross-product projection is close to the desired invariant Gibbs distribution, converging to it in the 
 $\varepsilon \to 0$ limit.  We apply similar ideas to those of  \cite{mattingly2010convergence} which consider invariant measures of numerical approximations of SDE solutions.  We start with Dynkin's Formula over one timestep of the SDE, and then replace the integral over the SDE solution with the MH solution, bounding the difference.  Summing over multiple timesteps and noticing a telescoping series, we show the long-time average over the MH solution converges to the average over the invariant measure of the SDE, which is the Gibbs distribution.
  Therefore, as in \cite{mattingly2010convergence}, we find that the difference between the invariant measures is the same order of magnitude as the error between the MH dynamics and the solution to the SDE \eqref{eq:colorSDE} on a finite time interval, given by \eqref{eq:36}.   
  
 Our goal is to show that the long-time average of a $C^\infty$ test function $\varphi$
   $$
\lim_{n\to\infty} \E[]{ \frac 1n \sum_{k=0}^{n-1} \varphi( \sig^k) }
 $$
 where  $ \sig^k $ is the $k$-th MH step with the inaccurate accept rate \eqref{eq:alpha} and any projection to form $\nui_i^n$ in \eqref{eq:proposalcolor}, converges to the stationary average $\bar{\varphi}$  with respect to the invariant measure $\mu$ of the  SDE \eqref{eq:colorSDE} with corresponding projection,
 \begin{equation}\label{barvar}
 \bar{\varphi} = \int \varphi(\sig) \mu( \sig) \mathrm{d} \sig.
 \end{equation}
  We build on the fact that the MH algorithm has a unique stationary distribution, that is not the Gibbs distribution, and that the SDE has a unique stationary measure $\mu$ because the generator $\mathcal{L}$ of the SDE  \eqref{eq:colorSDE} is hypoelliptic; its second order term is  
 \begin{align*}
 & \sum_{i,j} ( PC_N P^T )_{ij}\partial_i \partial_j = \\
 & \qquad (P^T D)^T C_N (P^T D)  - \sum_{i,j} \partial_i (PC_N P^T)_{ij} \partial_j,
 \end{align*}
 where $D$ is the diagonal matrix with $\bar{D}$ repeated 3 times along the diagonal and
 the system of vector fields $(P^T D)^T C_N (P^T D)$ covers $\mathbb{T}(\mathbb{S}^2)^N$ as in \cite{banas2014stochastic} and the second term on the right hand side is first order.  In the special case that the cross-product projection matrix $P_1$ in \eqref{eq:Pcross} is used, then the SDE has the known invariant measure of the Gibbs measure  $\mu$ in \eqref{eq:Gibbs}.  Our argument will therefore show that the MH algorithm with cross-product projection samples a distribution that converges to the Gibbs measure as the proposal size $\varepsilon\to0$.

We start with Dynkin's Formula \cite{oksendal2003stochastic} for the SDE \eqref{eq:colorSDE}, with generator $\mathcal{L}$, over a time-step $\delta t$, 
\begin{equation}\label{C1}\begin{aligned}
\E[]{\psi(\s ((k+1)\delta t))} - &\E[]{\psi(\s( k\delta t))} \\
& = \E[]{ \int_{k\delta t}^{(k+1)\delta t} \mathcal{L} \psi(\s(t)) \dt } .
\end{aligned}\end{equation}  
Consider that $\psi$ solves a Poisson equation for  $C^\infty$ test function $\varphi$,
\begin{equation}\label{C2}
\mathcal{L}\psi = \varphi - \bar{\varphi}
\end{equation}
where the stationary average $\bar\varphi$ is defined in \eqref{barvar}. Using \eqref{C2} in the right-hand-side of \eqref{C1}, we have that
\begin{equation}\label{C5}\begin{aligned}
\mathbb{E} \big[ \psi(\s ((k+1)&\delta t)) \big] -\E[]{\psi(\s( k\delta t))} \\
& = \E[]{ \int_{k\delta t}^{(k+1)\delta t}  \varphi(\s(t)) \dt } - \bar{\varphi} \delta t.
\end{aligned}\end{equation}
The integral term can be bounded by
\begin{equation}\label{C6}
\left| \E[]{ \int_{k\delta t}^{(k+1)\delta t}  \varphi(\s(t)) \dt  - \varphi(\s(k\delta t)) \delta t } \right| \le c \delta t^2
\end{equation}
for some constant $c$ independent of $\delta t$ by Riemann sum approximations of integrals.
 From Theorem \ref{thm:MHtoSDE}, the difference between the SDE solution $\s(k\delta t)$ and the Metropolis step $\sig^k$ is bounded by
 \[
 \E[]{ \left\| \sig^k - \s(k\delta t)  \right \| } \le c_3 \delta t^{1/4}, 
 \]
 and therefore for smooth test functions
 \begin{equation}\label{C8}
 \left| \E[]{\varphi( \s(k\delta t) ) }  - \E[]{\varphi(\sig^k) } \right| \le c_4 \delta t^{1/4}.
 \end{equation}
Using bounds \eqref{C6} and then \eqref{C8} we have that \eqref{C5} can written as 
\begin{align*}
    & \E[]{\psi(\s ((k+1)\delta t))} - \E[]{\psi(\s( k\delta t))} = \\
     &\hspace{1cm}  \E[]{\varphi(\sig^k)} \delta t - \bar{\varphi} \delta t + e_1
\end{align*}
where $|e_1| \le c \delta t \delta t ^{1/4}$ for some constant $c$ independent of $\delta t$.  Re-arranging, and dividing by $\delta t$ we have that
\begin{equation*}
 \E[]{\varphi(\sig^k)} - \bar{\varphi} = \frac{1}{\delta t}  \E[]{\psi(\s ((k+1)\delta t))-\psi( \s(k\delta t))} + e_2
\end{equation*}
where $|e_2| \le c  \delta t ^{1/4}$ for some constant $c$ independent of $\delta t$.   Summing over $n$ values of $k$ and dividing by $n$ we have that
\begin{equation}\begin{aligned}
\frac{1}{n}&\sum_{k=0}^{n-1} \E[]{\varphi(\sig^k)} - \bar{\varphi} \\
& =  \frac{1}{n\delta t} \sum_{k=0}^{n-1} \E[]{\psi(\s ((k+1)\delta t))-\psi( \s(k\delta t))} + e_2
\end{aligned}\end{equation}
which has a telescoping sum on the right-hand side.  By defining $T = n\delta t$ the above is equivalent to
\begin{equation}
\frac{1}{n}\sum_{k=0}^{n-1} \E[]{\varphi(\sig^k)} - \bar{\varphi} =  \frac{1}{T}  \E[]{\psi(\s (T)-\psi(\s( 0))} + e_2 .
\end{equation}
Recall that $\psi$ is the unique solution to the Poisson equation \eqref{C2} therefore it is smooth because $\varphi$ is smooth.  Indeed, the theory of hypoelliptic operators is precisely such that $\mathcal{L} u \in C^\infty$ implies $u \in C^\infty$, see \cite{hormander1967hypoelliptic} or \cite{hormander2015analysis}, Chapter $XI$.  Since we are operating on a compact space overall, $\psi$ is thus bounded and the convergence result follows.  Thus, the $1/T$ term goes to zero as $T\to\infty$ ($n\to\infty$).  We therefore conclude that the MH long-time average converges to the stationary average with respect to the SDE invariant measure $\bar{\varphi}$ as $\delta t\to 0$ ($\varepsilon\to 0$) with order $\delta t^{1/4}$ and as $n\to\infty$, and the following convergence results holds:

\begin{theorem} \label{thm:IM}
Define $\sig^n$ as the $n^{\textrm{th}}$ step of the MH dynamics with colored noise proposal given in \eqref{eq:proposalcolor}, either the cross- or cross-cross-projection,  accept rate given in \eqref{eq:alpha} and let $\mu(\s)$ be the invariant measure of the corresponding SDE \eqref{eq:colorSDE} with the same projection.  Then
$$
\left|  \frac{1}{n}\sum_{k=0}^{n-1} \E[]{\varphi(\sig^k)} - \int \varphi(\s) \mu(\s) \mathrm{d} \s \;  \right| \le \frac{c_1}{n \delta t} + c_2 \delta t ^ {1/4}
 $$
for time step $\delta t = \varepsilon^2 \beta / 2$ and constants $c_1$ and $c_2$ independent of $n$, and $\delta t$.
\end{theorem}

\begin{remark}
A nearly identical argument can be used to show that the invariant measure for the SDE with the $P_2$ projection will converge to that of the SDE with the $P_1$ projection as the colored noise converges to white noise, thus both SDEs sample the Gibbs measure.  In other words, as $\kappa \to 0$ the covariance matrix $C \to I$ in a uniform sense in our definition of \eqref{Cbar} as an operator on $\ell^2$ (and hence smooth) functions $(\mathbb{S}^2)^N$.
\end{remark}

\subsection{A New Non-local SPDE Limit\label{sec:PSDE}}

In this section, we discuss the extension of the non-local SPDE \eqref{eq:spde} to the case $\mathbb{T}^d \to \mathbb{S}^2$ with $d>2$ obtained by taking 
 the limit as $N\to\infty$ (with $\beta$ constant) of the SDE \eqref{eq:colorSDE} and remark briefly on properties of the corresponding solutions.
In particular, formally taking the limit of \eqref{eq:colorSDE}, we arrive at a non-local  stochastic version of the harmonic map heat flow equation given by
\begin{equation}\begin{aligned}
\label{eqn:spde}
d \sigma = &\left( -\sigma \times (M_\kappa (D)) (\sigma \times \Delta \sigma) \right) dt \\
&+ \sigma \times ( \mathcal{F}^{-1} (m(k))^{- \kappa} \circ dW (k)),
\end{aligned}\end{equation}
where we let $m( k) = 2 \pi k$ if $|k| \neq 0$, and $m( k) = 1$ for $k= 0$, $\mathcal{F}$ is the Fourier transform on $\bbT^d$, $M_\kappa (D)$ is the Fourier multiplier such that
\[
M_\kappa (D) f = \mathcal{F}^{-1}   (|m(k)|)^{-2 \kappa}  \mathcal{F} f  
\]
and $dW(k)$ are a set of independent standard Gaussian noises for each corresponding Fourier mode in frequency space. 
  Note that many other forms of the covariance structure could easily work here, such as $m(k) = \langle k \rangle = \sqrt{1 + |2 \pi k|^2}$.
Also note that we can write 
 \[
( ( I - \Delta)^{-\kappa} f)(x) = \int K_\kappa (x,y) f(y) dy
\]
with the integral kernel given by
\[
K_\kappa (x,y) = \frac{1}{(2 \pi)^d} \sum_{k \in \bbZ^d}  \int e^{-i 2 \pi k \cdot (x-y)} \langle k \rangle^{\kappa}. 
\]
Then, if $\kappa$ is chosen such that $M_\kappa (D)$ is trace class with a weight relating to the regularity required  ($\int K(x,y) dx , \int K(x,y) dy < \infty$ as well as integrals of derivatives of $K$), we can use canonical results on stochastic PDEs coupled with existence arguments for quasilinear heat equations.  We will follow somewhat the ideas in \cite{de1999stochastic,gess2016stability} for stochastic PDEs with multiplicative noise (mostly in the context of motivating the It\^o formulation in the former and for using energy estimates to handle degenerate SPDE models in the latter).  For the key energy estimates on the deterministic piece, we cite the general theory of well-posedness for quasilinear heat equations developed in \cite[Chapter $15$]{Tay3}.  For possible extensions to non-trace class covariance structure, see the recent work of \cite{bruned2019geometric} where a renormalization is proposed.  It will be a topic of further work to explore the place of our colored noise model within this context.  

Using the regularity of the colored noise, we provide a brief outline of existence for solutions to \eqref{eqn:spde} in Appendix \ref{A:LWP}.  However, as the results are fairly standard with sufficiently regular noise, we proceed with a detailed numerical study of convergence of the Metropolis-Hastings model and dynamics.

\begin{figure*}
	\includegraphics[width=\textwidth]{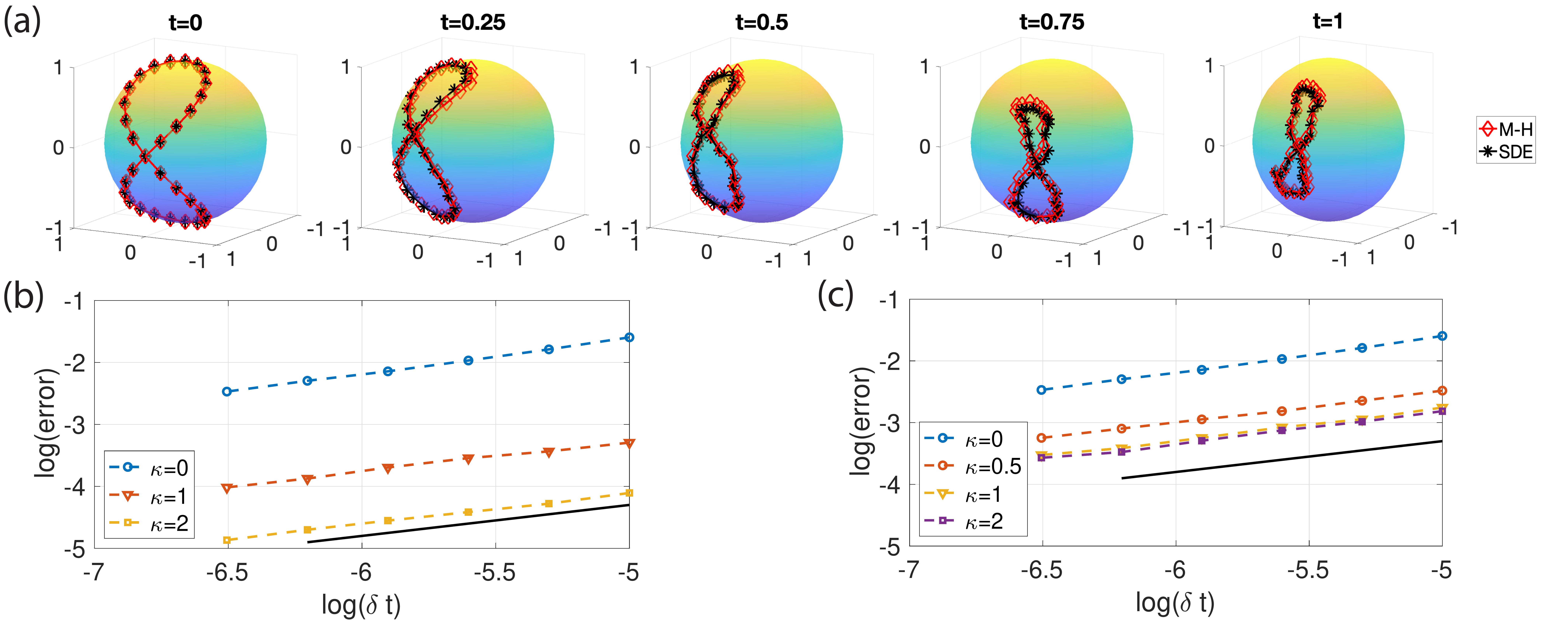}
	\caption{(a) Dynamics of MH algorithm and SDE \eqref{eq:colorSDE} at the indicated values of time, $t$ (recall the relationship between SDE time-step and MH proposal size: $\delta t = 2\beta \varepsilon^2$). Parameters: number of spins $N=32$, inverse temperature $\beta = 10$, the order of eigenvalues $\kappa = 1$, time step size $\delta t = 1e{}^-5$.
(b) 	Strong order of convergence with respect to time step size for the error between MH algorithm and SDE \eqref{eq:colorSDE}, both using the cross projection matrix \eqref{eq:Pcross}.  The solid black line with slope 1/2 indicates the order is approximately that given in Theorem \ref{thm:MHtoSDE}.  (c) The same as (b) but using the cross-cross projection matrix \eqref{eq:Pcrosscross}. (b) and (c) parameters: $N=16$, $\beta=5$, error calculated at time $T=0.05$, and averaged over 400 simulations. 
	}
	\label{fig:mh_sde}
\end{figure*}

\section{Numerical Results}
\label{sec:num}

In this section, we perform numerical simulations to support our convergence results and demonstrate the discussed differences when using different projections.  All the simulations are from the one dimensional periodic lattice $\mathbb{T}^1 $ to the unit sphere $\mathbb{S}^2$.
The MH dynamics are simulated as explained in Sec.~\ref{sec:colored_noise}. To numerically solve the SDE \eqref{eq:colorSDE_ito}, written in the It\^o form, we use the stochastic Euler's method combined with a normalizing step to project the spins back onto the sphere after each time step.

We start by showing a trajectory-wise comparison in  Figure \ref{fig:mh_sde}(a) of the MH dynamics and the SDE dynamics generated utilizing the same random noise for the proposal in the MH as the diffusion term in the SDE.  Each spin is plotted on the same sphere, with lines connecting nearest neighbors.
Figures \ref{fig:mh_sde}(b) and (c) show the strong order of convergence for the error between the MH algorithm and the SDE \eqref{eq:colorSDE} with respect to the time step size $\delta t$, for which the equivalent MH proposal size is  $\varepsilon = \sqrt{2 \delta t / \beta} $.  The error is calculated at fixed time $T$ as
\begin{equation}
\mathbb{E}\left[  \frac{1}{N}\sum_{i=1}^N \| \sigi_i^n - \si_i(n\delta t) \|^2  \;\right],
\end{equation} 
where the expectation is taken over multiple realizations. The numerical convergence order is approximately $\frac 12$, supporting Theorem \ref{thm:MHtoSDE} as a tight bound on the error regardless of choosing the cross projection matrix \eqref{eq:Pcross} or the cross-cross projection matrix \eqref{eq:Pcrosscross}.

\begin{figure}[ht!]
	\includegraphics[width=0.42\textwidth]{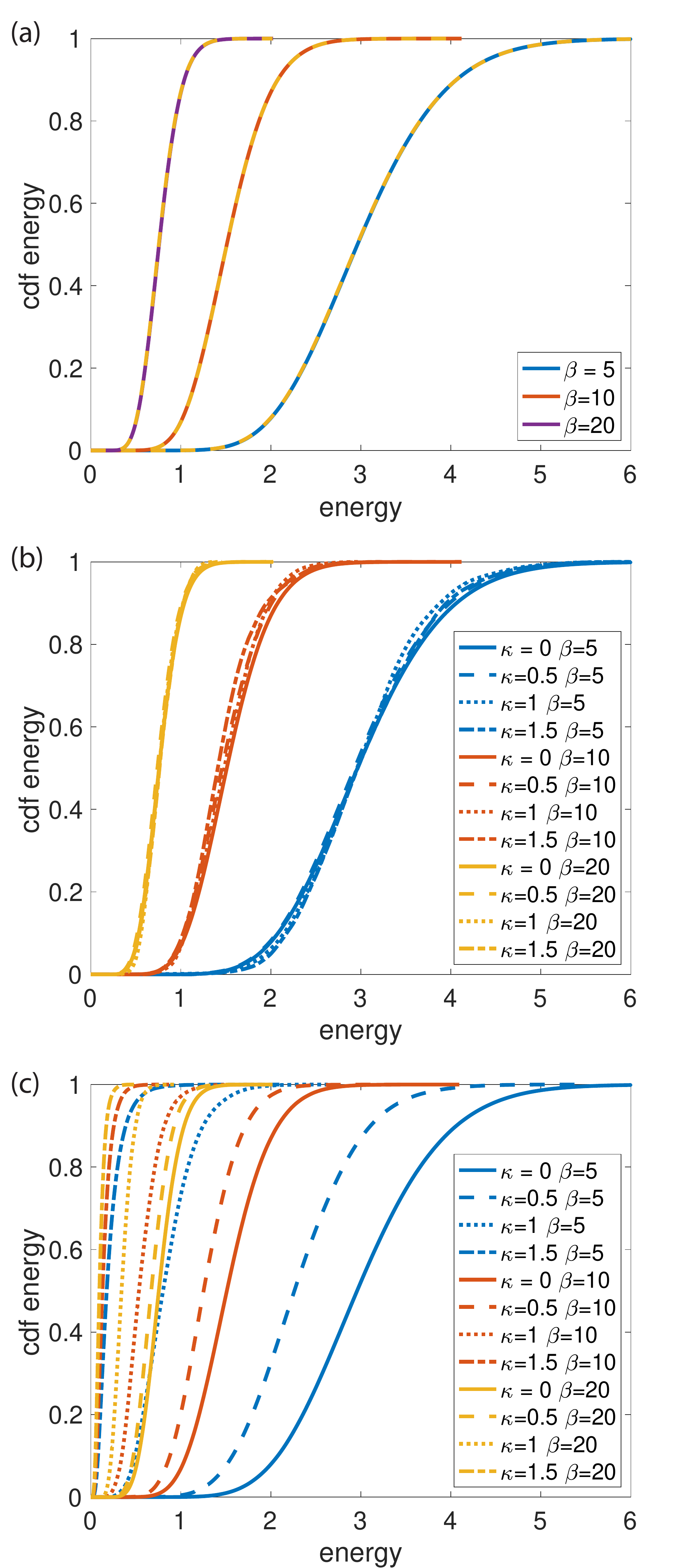}
\caption{ (a) Energy distribution in equilibrium for the indicated values of $\beta$ when ``white noise'' $\kappa=0$ is used.  The solid lines are for the cross-product projection, the yellow dashed lines are for the cross-cross-product projection.  They agree entirely; Gibbs is being sampled in all cases.  (b) For the cross-product, at the indicated values of $\kappa$, the same distribution is being sampled at each of the three values of $\beta=5,10$ and $20$.  (c) For the cross-cross-product, different distributions are being sampled for different values of $\kappa$, consistent with Gibbs not being the invariant measure of the SDE when $\kappa\ne 0$.  
Parameters: $N=16$, $\delta t = 1e-4$, and $1e5$ different time points. }
	\label{fig:distribution}
\end{figure}

Next we show the effect of the different projection matrices on the invariant measure of the SDE system \eqref{eq:colorSDE}.  Since the desired invariant measure is high-dimensional, we instead plot the empirical cumulative distribution function (cdf) of the energy over time.    Figure \ref{fig:distribution}(a) shows that for the case of white noise, $\kappa = 0$, utilizing either the cross projection matrix \eqref{eq:Pcross} or the cross-cross projection matrix \eqref{eq:Pcrosscross} results in indistinguishable invariant distributions of the energy; both versions have the Gibbs distribution as an invariant measure.  However, when coloring the noise by increasing $\kappa$, it is only the cross projection matrix \eqref{eq:Pcross} that maintains an energy distribution indistinguishable from the white noise case.  Figure \ref{fig:distribution}(b) supports that the color noise SDE \eqref{eq:colorSDE} with the cross projection matrix \eqref{eq:Pcross} is ergodic with respect to the correct Gibbs distribution, despite being the limit of our incorrect MH scheme in Sec.~\ref{sec:colored_noise}. 
Figure \ref{fig:distribution}(c) shows that the SDE system with the cross-cross projection matrix \eqref{eq:Pcrosscross} has lower energy on average as the correlations in the colored noise increase with increasing $\kappa$.

To further illuminate this interaction of the projection matrix and the correlated noise, we look how each term in the SDE effects the energy of system when in equilibrium.  The energy, $H$ given by \eqref{eq:H}, evolves according to the It\^o SDE
\begin{equation}\begin{aligned}\label{eq:dE_color}
dH =& \frac1N \sum_{i=1}^N N^2 (\si_{i+1}-\si_i) \cdot ( d\si_{i+1} - d\si_i ) \\
& + N\beta^{-1} \Tr( C_N^{1/2}P^T A P C_{N}^{1/2} )dt,
\end{aligned}\end{equation}
where $A$ is the tri-diagonal matrix with 2 on the diagonal and -1 on the sub- and super-diagonals (taking into account periodicity), $d\si_i$ is given by \eqref{eq:colorSDE_ito}, and $\si_{N+1}=\si_{1}$.
Note that since $\Tr(XY) =\Tr(YX)$ for two $n\times n$ matrices $X$ and $Y$ , the trace term 
\begin{align*} 
& \Tr(PC_N^{1/2}C_N^{1/2}P^TA) = \Tr(PP^T C_N A) \\ & = \Tr(PP^T\phi \bar{D}^2 \phi^T A) 
 = \Tr(PP^T \phi \phi^T \bar{D}^2 A) \\ & = \Tr(P \bar{D}^2 A)  = 4 N \sum_{i=1}^N \lambda_i^2
\end{align*}
is a constant independent of the choice of projection matrix.
We therefore ignore this term and proceed to decompose $d\si_i$ given by \eqref{eq:colorSDE_ito} over one $\delta t$ time-step of numerical integration as
\begin{equation}\label{eq:ds_pq}
\si_i^{n+1}-\si_i^n = \boldsymbol p_i^n \delta t - 2\beta^{-1}\frac{\Tr(\bar{C}_N)}{N} \si_i^n \delta t + \boldsymbol q_i^n \sqrt{\delta t},
\end{equation}
where we define 
\[ 
\vec p^{\;n} = P \frac{1}{N} C_N P^T \Delta_N \vec s^{\;n}  \textrm{ and } \vec q^{\;n} = PC_N^{1/2}\vec w^{\;n}
\] as well as take 
$\si^n_{N+1} = \si^n_1$, $ \boldsymbol p^n_{N+1} =  \boldsymbol p^n_{1} $, and $ \boldsymbol q^n_{N+1} =  \boldsymbol q^n_{1} $ for the periodic boundary conditions.  The trace term in \eqref{eq:ds_pq} is also of a form independent of the choice of projection matrix.  Therefore, to illuminate the interaction of the projection matrix and the correlated noise we consider only the contributions to \eqref{eq:dE_color}, the change in energy, given by the $\vec p^{\;n}$ and $ \vec q^{\;n}$ terms over each time-step of the numerical integration of the SDE, calculated as 
\begin{equation}\begin{aligned}\label{deltaE_drift}
\delta H^n_{\textrm{drift}} &=   \sum_{i=1}^N (\si^n_{i+1}-\si^n_{i})\cdot ( \boldsymbol p^n_{i+1} - \boldsymbol p^n_i) \delta t 
\end{aligned}\end{equation}
and 
\begin{equation}\begin{aligned}\label{deltaE_noise}
\delta H^n_{\textrm{noise}} &= N \sqrt{\frac{2}{\beta}} \sum_{i=1}^N (\si^n_{i+1}-\si^n_{i})\cdot ( \boldsymbol q^n_{i+1} - \boldsymbol q^n_i) \sqrt{\delta t } .
\end{aligned}\end{equation}

\begin{figure*}[t]
\includegraphics[width=\textwidth]{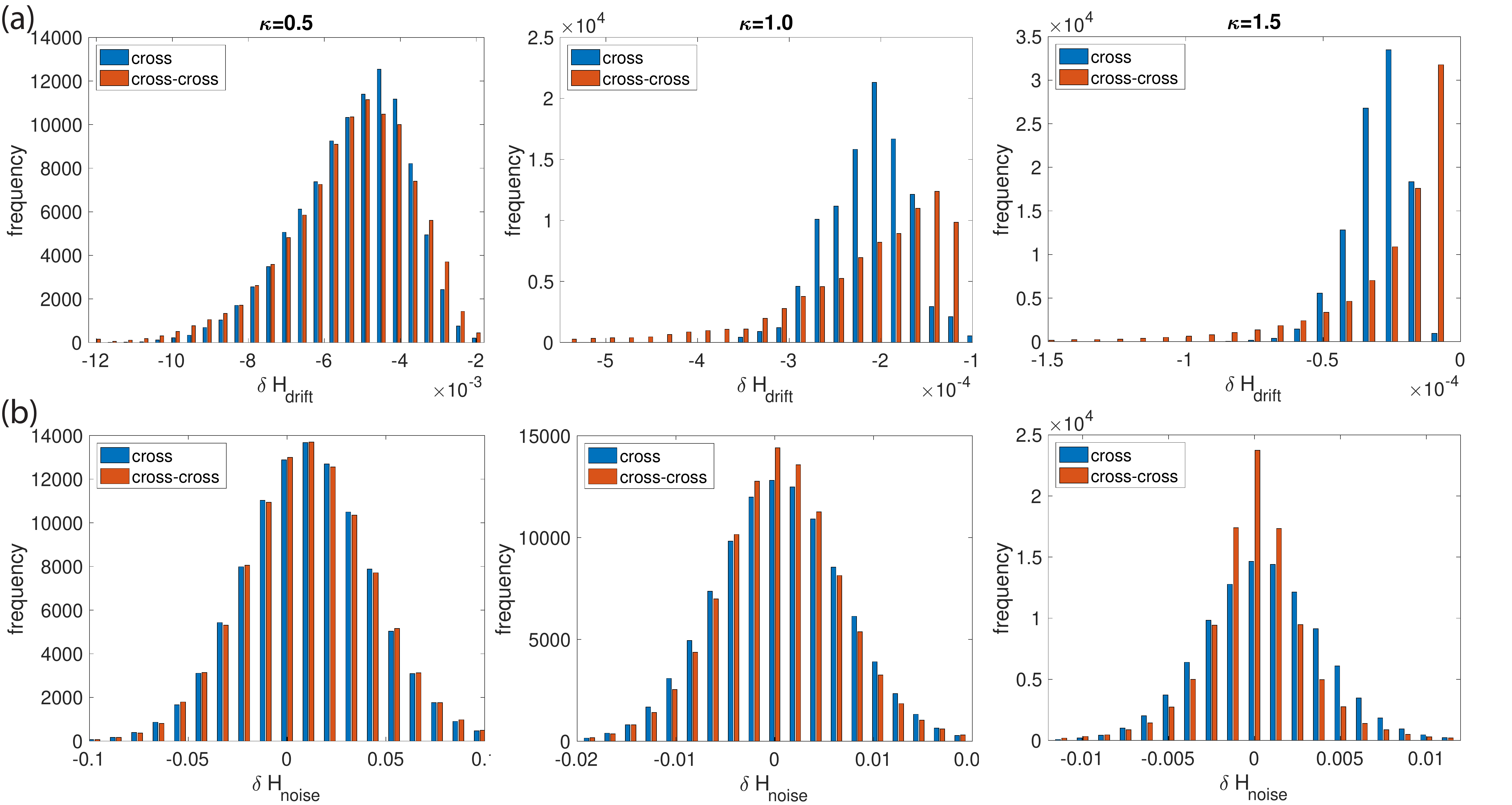}
\caption{Comparison of the effect on the energy for the cross and the cross-cross projection matrix in SDE \eqref{eq:colorSDE}. (a) The empirical histogram of $\delta H^n_{\textrm{drift}}$ from \eqref{deltaE_drift} taken at each time point of one simulation of the SDE in equilibrium for the indicated value of $\kappa$. (b) Same as (a) but for $\delta H^n_{\textrm{noise}}$ from \eqref{deltaE_noise}. 
Parameters:  $N=16$, $\delta t = 1e-4$, $\beta = 10$, and $1e5$ different time points.} 
	\label{fig:deltaE}
\end{figure*}

In Fig.~\ref{fig:deltaE} we plot the distribution of $\delta H^n_{\textrm{drift}} $ and $\delta H^n_{\textrm{noise}}$ for both the $P_1$ (cross-product) and $P_2$ (cross-cross-product) projections 
over the course of one simulation using each of the indicated values of $\kappa$ to form $\bar{C}_N$.  We see that as $\kappa$ increases, the differences between these distributions increases, consistent with Fig.~\ref{fig:distribution}(c) showing more deviation from the Gibbs distribution with increasing $\kappa$.  This difference is more pronounced in the deterministic drift contribution to the energy, $\delta H^n_{\textrm{drift}}$, than the diffusion contribution, $\delta H^n_{\textrm{noise}}$.  It suggests the random-walk nature of the dynamics remains relatively unaffected by the choice of projection, while the cross-cross projection produces  long tails to lower values of $\delta H^n_{\textrm{drift}}$ possibly explaining the shift in average energy to lower energies seen in Fig.~\ref{fig:distribution}(c).

\begin{figure}[ht!] 
\includegraphics[width = 0.4\textwidth]{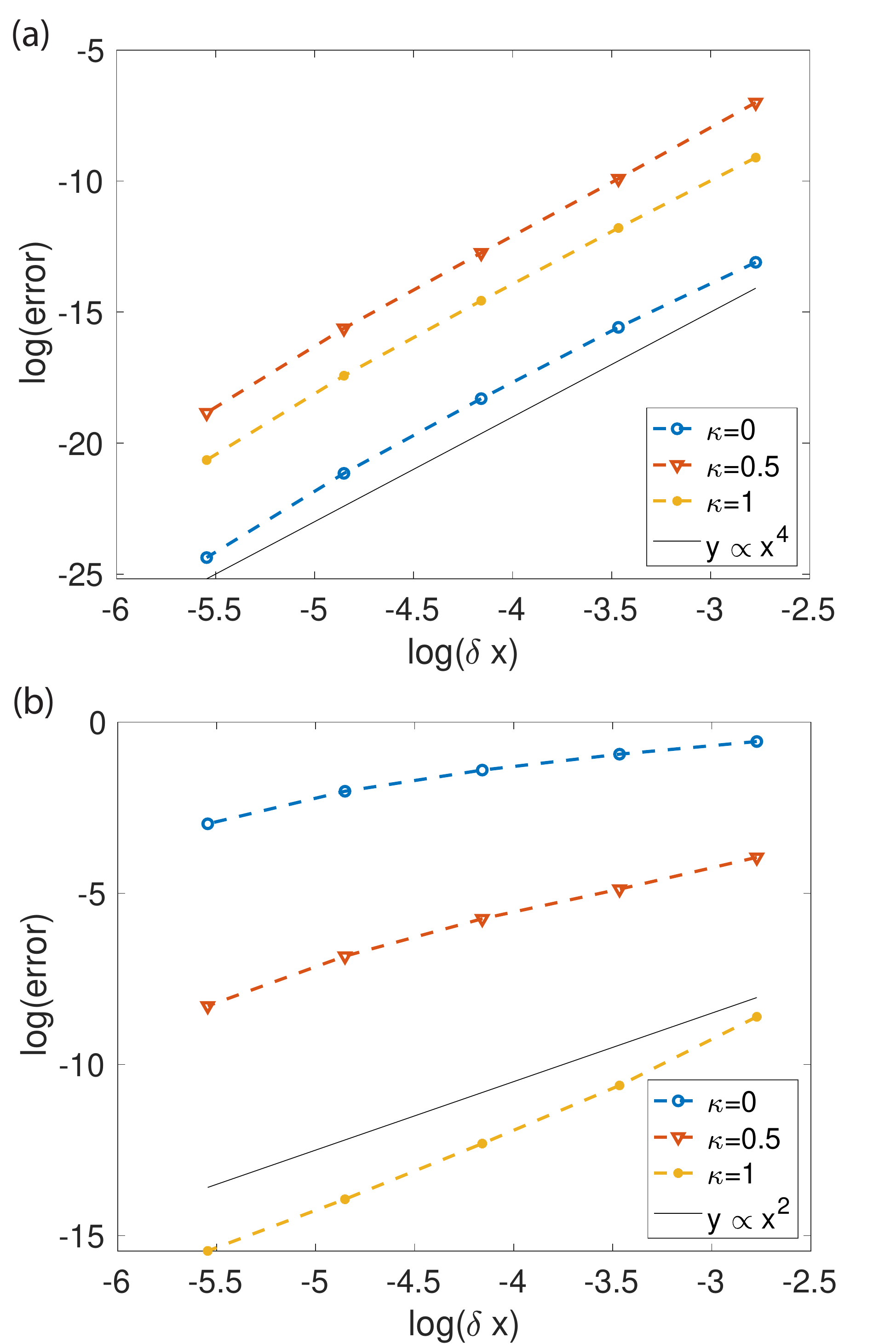}
 \caption{(a) Convergence plot for the deterministic finite difference approximation of the non-local PDE \eqref{eq:colorSDE} with $\beta^{-1}=0$.  (b) Convergence plot for the Stochastic finite difference approximation of the non-local SPDE \eqref{eq:colorSDE} with $\beta=5$, averaged over 100 simulations.  In both panels, dynamics are simulated until $T=0.125$ with $\delta t = \tfrac12 \delta x^2$. }
 \label{fig:sdeconvplot}
\end{figure}

In Fig.~\ref{fig:sdeconvplot} we verify convergence of the SDE system to the SPDE \eqref{eq:spde}.  First, in Fig.~\ref{fig:sdeconvplot}(a), for just the deterministic drift part of this system, we show convergence of the finite difference ODE approximation of the non-local PDE ($\beta^{-1}=0$).  We compute the error at fixed time $T$ between each coarser scale, $N=N_c$, with the finest scale, $N=N_f$, as
\begin{equation}
 \frac{1}{N_c}\sum_{i=1}^{N_c} \| \si_i^{\textrm{Coarse}}(T) - \si_{1 + (i-1)\frac{N_f}{N_c}}^{\textrm{Fine}}(T) \|^2  .
\end{equation}
Then, in Fig.~\ref{fig:sdeconvplot}(b) we shown the strong convergence of the SDE, taking the expectation of the above error over realizations.  Note the convergence rate even for the white noise case of $\kappa=0$, which is not guaranteed if more than one spatial dimension of this SPDE was considered due to the potential breakdown of regularity of the deterministic solution in that case.  The deterministic convergence of order 4 is twice that of the noisy system, which is approximately order $2$.

\begin{figure*}[ht!]
	\includegraphics[width=\textwidth]{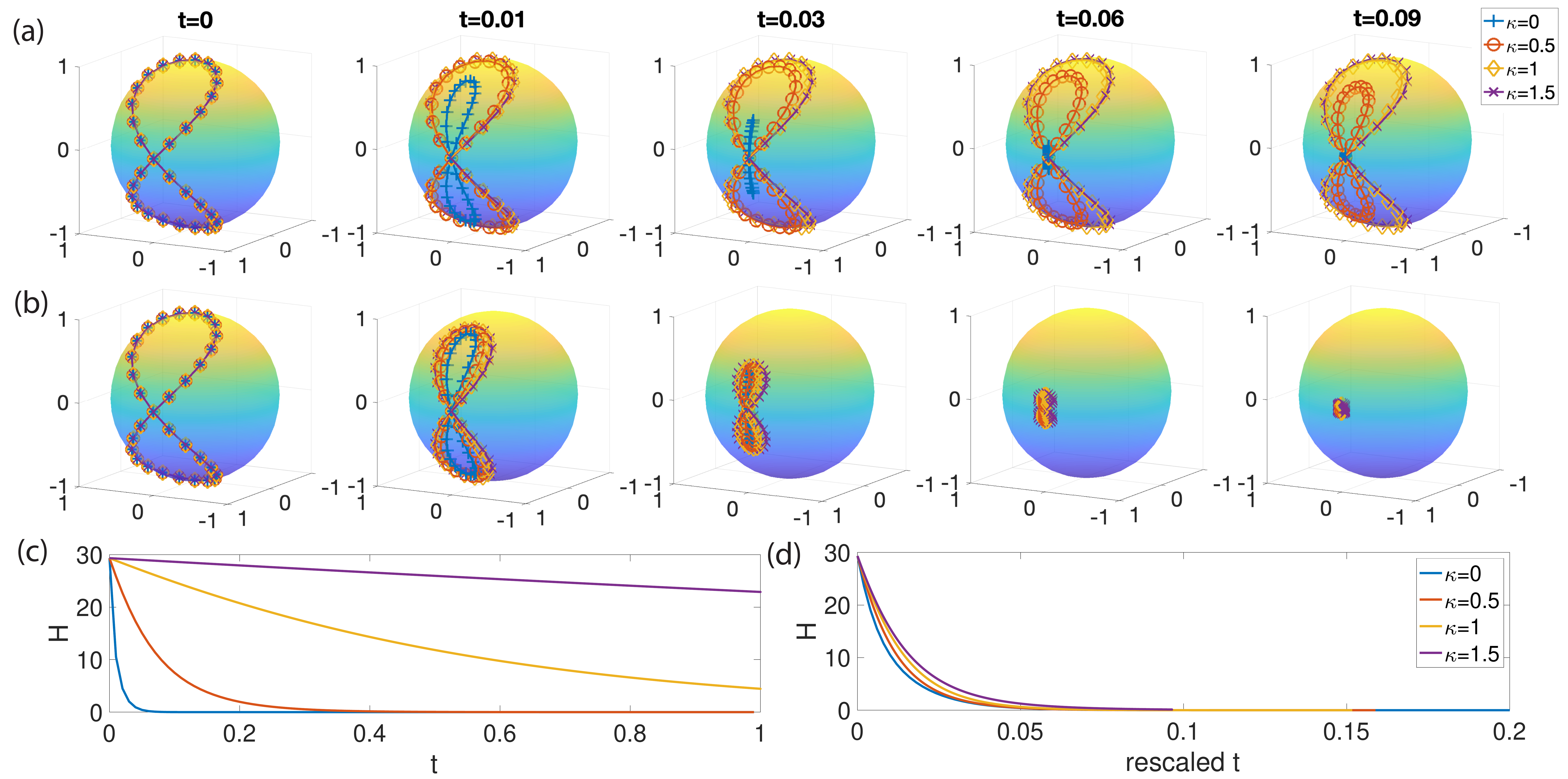}

	\caption{(a) Dynamics of the non-local PDE (Eq.~\ref{eq:spde} with $\beta^{-1}=0$) at the indicated values of time, $t$, with original timescale and (b) with rescaled time $\tilde{t} = (2\pi)^{2 \kappa} t$ for $\kappa \in \{0,0.5,1,1.5 \}$.  (c) Evolution of the energy with the original timescale and (d) with rescaled time. 
	Legend applies to all panels.  Parameters: number of spins $N=32$, original time step size $\delta t = 1e{}^-5$.  
	}
	\label{fig:PDEdynamics}
\end{figure*}

Last, we look at some of the behavior of the new non-local (deterministic) PDE.  In Fig.~\ref{fig:PDEdynamics}(a) we show the evolution toward equilibrium of the spins for different values of $\kappa$ highlighting the different time scales.  By considering the covariance operator as a fractional Laplacian, acting similarly to the harmonic map heat flow equation, we conjecture the time rescaling being related to the diffusion time scaling of the underlying non-local heat equation
\[
u_t = M_\kappa (D) \Delta u,
\]
which decays to its equilibrium on the time scale $e^{-\lambda_1^{2(1-\kappa)}}$ for $\lambda_1$ the first non-trivial eigenvalue of the Laplacian on $\bbT^d$.  The non-local form of the operator we consider here does not immediately present a leading order linear operator of this form as occurs in the cross-cross projection, however we will see that this time scale still arises in  Figure~\ref{fig:PDEdynamics}.   Figures~\ref{fig:PDEdynamics}(c) and (d) also shows the effect of this time rescaling when looking at the evolution of the energy of the system.

\section{Conclusions/Discussions\label{sec:conclusions}}

We establish here a new stochastic partial differential equation as the limit of a set of sampling algorithms where the proposal is taken with spatially correlated colored noise, thereby deriving a mesoscopic model of fluctuations for spin systems in a principled way. 
The geometric nature of our system means that the nonlocal form of the drift arises in a manner that we have not seen before in the literature.  In order to ensure that the system samples the desired Gibbs measure, we have to be careful with the manner by which we project the noise into the geometric setting.  Specifically, we show using a cross-product projection samples the Gibbs measure while a cross-cross-product projection samples an invariant measure that is shifted to lower energy than the Gibbs measure.  This shift increases as the correlation length-scale of the noise is increased, and is shown numerically to be related to the deterministic effect on the energy of the system rather than fluctuations in the energy.  In addition to finding convergence rates, numerical simulations are also used to show that the nonlocal drift term of the new SPDE exhibits the same time-scales for relaxation to equilibrium as a fractional Laplacian, thereby acting similarly to the harmonic map heat flow equation.
Future work will involve considering other geometries beyond the sphere, performing a more careful analysis of the resulting SDE/SPDE systems following for instance the recent developments on geometric renormalization tools in \cite{bruned2019geometric}.

\appendix

\section{Convergence Proof of MH to SDE\label{app:SDEderivation}}

This Appendix compliments Sec.~\ref{sec:MHtoSDE} in the main text, containing the details of the derivation and proof of Theorem \ref{thm:MHtoSDE}.  In Sec.~\ref{App:OneStep} we derive the expansion of the mean drift of one MH step, \eqref{eq:4}, and bound the remainder terms.  In Sec.~\ref{App:Diff} we bound the error of the diffusion approximation of one MH step, \eqref{eq:diff_approx}.  Then in Sec.~\ref{App:Existence} we argue the existence and uniqueness of the solution to SDE \eqref{eq:colorSDE} and complete the proof of Theorem \ref{thm:MHtoSDE} giving the ideas for bounding the error between the MH and SDE dynamics.  Throughout this appendix we use $c$ or $c(\cdot)$ to indicate a positive constant, potentially different constant every time the symbol is used, independent of the parameters we are bounding quantities in, either $\varepsilon$ or $\delta t$.  The functional representation $c(\cdot)$ indicates precisely which model parameters the constant is dependent on.

\subsection{Expectation of One Metropolis Hastings Step \label{App:OneStep}}

In this section, we calculate the leading order in $\varepsilon$ terms of the expectation of one MH step, Eq.~\ref{eq:4} in the main text, and bound the expectation of the error of each expansion and approximation we utilize.  The leading terms which determine the dynamics are of size $\varepsilon^2$ while the mean-squared error is bounded by the next order terms of size $\varepsilon^6$. For brevity, throughout this section we drop the superscript $n$ on all terms.  The first error term, denoted $\vec{r}_1$ and defined coordinatewise as $(\boldsymbol{r_1})_i \in \mathbb{R}^3$, $i=1\dots N$, arises from utilizing the expansion \eqref{eq:expansion} of the MH proposal as follows
\begin{equation}\label{r1i}
(\boldsymbol{r_1})_i = \E{ \boldsymbol{d}_i (1 \wedge e^{-\beta \big( H(\sigp{}) - H(\sig) \big) }) },
\end{equation}
where 
\begin{dmath}\label{di}
 \boldsymbol{d}_i \equiv \sigpi_i - ( \sigi_i + \varepsilon \nui_i - \frac{\varepsilon^2}{2} \| \nui_i\|^2 \sigi_i)
\end{dmath}
is the difference between the true proposal and its approximation.  Other error terms will come from computing the expectations of the two terms of the expansion of the proposal. Namely, by writing 
\begin{equation}\begin{aligned}
\label{A1}
 \EE&\left[(\sigpi_i - \sigi_i) (1 \wedge e^{-\beta \big( H(\sigp{}) - H(\sig) \big) })\right] \\
 &= \E{ \varepsilon \nui_i (1 \wedge e^{-\beta \big( H(\sigp{}) - H(\sig) \big) })} \\
 &- \E{ \frac 12 \varepsilon^2 \|\nui_i\|^2 \sigi_i  (1 \wedge e^{-\beta \big( H(\sigp{}) - H(\sig) \big) })} +  (\boldsymbol{r_1})_i,
\end{aligned}\end{equation}
we now can move to defining the error term denoted $\vec r_2$, which bounds the difference of the expectations of the $\nui_i$ term above when using the actual difference in the Hamiltonians of the proposal and current step verses its leading order approximation \eqref{deltaH}.  Namely, we have
\begin{dmath}\label{R2b}
\vec{r}_2 =  \E{ \varepsilon P C_N^{1/2} \w (1 \wedge e^{-\beta \big( H(\sigp{}) - H(\sigma) \big) })} \\
-  \E{ \varepsilon P C_N^{1/2} \w (1 \wedge e^{-\beta \varepsilon (P C_N^{1/2} \w)^T \nabla H })}.
\end{dmath}
This error term $\vec{r}_2$ is bound in Sec.~\ref{App:Drift}.   We further bound the error of the approximation \eqref{eq:2} in the main text by bounding the term denoted by $\vec{r}_3$ 
\begin{align}
& \E{ \varepsilon P C_N^{1/2} \w (1 \wedge e^{-\beta \delta H })} \notag \\
& \hspace{1cm} = -\frac12 \varepsilon^2 \beta PC_NP^T \nabla H  +  \vec{r}_3 \label{r3r4def}
\end{align}
to be specifically defined later but that results from approximations applying Lemma \ref{lm:mps}.
A similar bound on the second, order $\varepsilon^2$, term in \eqref{A1} is handled in Sec.~\ref{append:ito_correction}, defining
\begin{dmath}\label{err_r5i}
( \boldsymbol{r_{4}} )_i \equiv \E{- \frac 12 \varepsilon^2 \|\nui_i\|^2 \sigi_i (1 \wedge e^{-\beta \big( H(\sigp{}) - H(\sig) \big) })} - \E{ - \frac 12 \varepsilon^2 \|\nui_i\|^2 \sigi_i  }.
\end{dmath}
 We also show in Sec.~\ref{append:ito_correction} that this term is the \Ito correction term to the drift between the Stratonovich and \Ito representations of the SDE, Eqs.~\eqref{eq:colorSDE} and \eqref{ito} respectively. 
 
Combining the to be presented error bounds in Eqs. \eqref{R1}, \eqref{R2}, \eqref{R4}, \eqref{R5} below, we arrive at the global bound of Eq.~\ref{eq:4} in the main text, namely by defining
\begin{multline*}
 \E{ (\tilde{\sigma^n} - \sigma^n) (1 \wedge e^{-\beta \big( H(\tilde{\sigma^n}) - H(\sigma^n) \big) })} \\ 
 = - \frac 12 \beta \varepsilon^2 P C_N P^T \nabla H -\varepsilon^2 \frac 1N \Tr(\bar{C}_N) \sigma + \sum_{k=1}^4 \vec{r}_k,
\end{multline*}
 the error is bounded by
\begin{dmath}\label{Rn}
 \E[]{\| \sum_{k=1}^4 \vec{r}_k  \|^2} \le c N^{13} \beta^6 (\Tr(C_N))^3 \varepsilon^6,
\end{dmath}
where we will repeatedly use the very crude bound 
\[
\| \nabla H \| \lesssim N ^2
\]
from \eqref{gradHdef}.   

\begin{remark}
Intuitively the term $\nabla H$ should have a better bound in some probability sense, but the current bound might be the best we could hope for so far since we are not able to specify a probability distribution for $H$ with the wrong MH step setting.
\end{remark}


\subsubsection{Bounding the Proposal Expansion Remainder\label{App:Prop_remainder}}

In this section we bound the error of the remainder terms, $(\boldsymbol{r_1})_i$, defined in \eqref{r1i}.  

We begin by bounding $\boldsymbol{d}_i$ in \eqref{di} by first bounding the size of the tangent vector $\vec\nu = P C_N^{1/2} \w$.
Since  the projection matrix $P$ acts either as the cross product, or the cross-cross product, the magnitude $\| Px \|$ is smaller than $\|x\|$ for each single spin vector, and we have that
\begin{dmath}
\| \vec\nu \|^2 = \| P C_N^{1/2} \w \|^2 \le \| C_N^{1/2} \w \|^2  =  (\w)^T C_N \w.
\end{dmath}
The expectation of $(\w)^T C_N \w$ is equivalent to the expectation of $N \sum_{q\in\{x,y,z\}} (\wq_q)^T \bar{D}^2 \wq_q$ with the way we have defined $\bar{C}_N$ in \eqref{Cbar}.    Since each component of $w$ is independent and identically distributed,
$$
\mathbb{E}\left[ \sum_{q\in\{x,y,z\}} (\wq_q)^T \bar{D}^2 \wq_q \right] = \mathbb{E}\left[ 3  \sum_{i=1}^N \lambda_i^2 w_{ix}^2 \right].
$$

 Bounding the expectation of $\|\vec\nu\|^{2k}$ for any positive integer $k$, we have that
\begin{dmath*}
 \E[]{ \| \vec\nu \|^{2k} } \le \E[]{\left( 3 N \sum_{i=1}^N \lambda_i^2 w_{i,x}^{2} \right)^k} \\
 = \E[]{ \sum_{i=1}^N N^k 3^k \lambda_i^{2k} w_{i,x}^{2k} + \textrm{cross terms with } w_{i,x}^{2l} w_{j,x}^{2(k-l)}} \\
 \le \E[]{w_{i,x}^{2k}} \left(3N \sum_{i=1}^N \lambda_i^2 \right)^k
 =  \left[ (2k-1){!}{!} \right] \left(3N \sum_{i=1}^N \lambda_i^2 \right)^k
\end{dmath*}
since $\E[]{ w_{i,x}^{2l} w_{j,x}^{2(k-l)}} = (2l-1)!! (2k-2l-1)!! \le \E[]{w_{i,x}^{2k}} = (2k-1)!! $.
 Therefore
\begin{dmath} \label{eq:bound_nu}
 \E[]{\| \vec \nu \|^{2k}} \le  \left[ (2k-1){!}{!} \right] \left( 3 \sum_{i=1}^N N \lambda_i^2 \right)^k  \\ 
 \le  c  \left( \Tr(\bar{C}_N ) \right)^k .
\end{dmath}

Now $\boldsymbol d_i$ can be bounded as a function of $\nui_i$ and the remainder from the Taylor expansion of 
$f(x) = (1+x)^{-1/2} = 1 - \frac x2 + \frac38 (1+\xi)^{-5/2}x^2 $ for some $\xi \in [0,\infty)$. 
Writing
$$
(1 +\varepsilon^2 \|\nui_i^n \|^2)^{-1/2} = 1 -
\frac{\varepsilon^2}{2} \| \nui_i^n \|^2 +  \boldsymbol \eta_i^n 
$$
we see that
\[
\| \boldsymbol \eta_i^n \|  \le \frac 38 \varepsilon^4 \| \nui_i^n \|^4
\]
and
\begin{equation}
\boldsymbol d_i^n = -\frac{\varepsilon^3}{2} \|\nui_i^n\|^2 \nui_i^n + (\sigma_i^n + \varepsilon \nui_i^n)  \boldsymbol \eta_i^n .
\end{equation}
We bound
$$\begin{aligned}
 \sum_{i=1}^N \E[]{\| \boldsymbol d_i \|^{2k}} 
 \le &\sum_{i=1}^N 4^k \E[]{\left\| -\frac{\varepsilon^3}{2} \| \nui_i \|^2 \nui_i \right\|^{2k}} \\
& + 4^k \E[]{\| (\sigi_i + \varepsilon \nui_i^n) \boldsymbol \eta_i \|^{2k}} ,
 \end{aligned}$$
then bound the first term
 $$
 \sum_{i=1}^N 4^k \E[]{\left\| -\frac{\varepsilon^3}{2} \| \nui_i \|^2 \nui_i \right\|^{2k}}  \le \sum_{i=1}^N \varepsilon^{6k} \E[]{\| \nui_i \|^{6k}}  
 $$
 and the second term
 $$\begin{aligned}
4^k \mathbb{E}&\left[\| (\sigi_i + \varepsilon \nui_i^n) \boldsymbol \eta_i \|^{2k}\right] \\
&\le 4^k \left( \E[]{\| \sigi_i + \varepsilon \nui_i \|^{4k}} \right)^{1/2}  \left( \E[]{ (\boldsymbol \eta_i)^{4k}} \right)^{1/2} 
\end{aligned}$$
to see that
$$\begin{aligned}
 \sum_{i=1}^N \E[]{\| \boldsymbol{d}_i \|^{2k}}  \le & \sum_{i=1}^N \varepsilon^{6k} E[ \| \nui_i \|^{6k} ]  \\
& + 4^k \left( \E[]{4^k + 4^k \varepsilon^{4k} \| \nui_i \|^{4k}} \right)^{1/2} \\
& \times \left( (\frac 38)^{4k} \varepsilon^{16k} \E[]{\| \nui_i \|^{16k}} \right)^{1/2} \\
 \le & c(k) [ \Tr(C_N) ]^{3k} \varepsilon^{6k}.
\end{aligned}$$

Together with the fact that 
 $\left(1 \wedge e^{-\beta \big( H(\sigp{}) - H(\sigma) \big) }\right) \le 1$ we have that
\begin{dmath}\label{R1}
 \E[]{ \| \vec r_1 \|^2 } \le \sum_{i=1}^N \E[]{ \| \boldsymbol{d}_i \|^2}  \le c N [ \Tr(C_N) ]^{3} \varepsilon^6 .
\end{dmath}

\subsubsection{The Drift Term\label{App:Drift}}

Here we bound $\vec{r}_2$ defined in \eqref{R2b}, bounding the error of approximating the change in the Hamiltonian between the current step and the proposal as in Eq.~\eqref{deltaH}.  This further allows us to establish the leading order approximation \eqref{w_lemma2}.  We bound the errors $\vec{r}_3$ and $\vec{r_4}$


We start with the approximation \eqref{deltaH}, a combination of a Taylor expansion in $\varepsilon$ and the approximation of $\sigp{} \approx \sig + \varepsilon \vec \nu$, the error of which is 
\begin{dmath} \label{eq:g}
 g \equiv \sum_{i=1}^N \frac{\partial H}{\partial \sigi_i} \cdot \boldsymbol{f}_i  + N \sum_{i=1}^N (\sigpi_i - \sigi_i) \cdot (\sigpi_i - \sigi_i) \\
 - \frac 12 N \sum_{i=1}^N (\sigpi_i - \sigi_i) \cdot (\sigpi_{i+1} - \sigi_{i+1} +\sigpi_{i-1} - \sigi_{i-1} ),
\end{dmath}
where the error $\boldsymbol{f}_i$ is from the first order Taylor expansion of the proposal,
\begin{equation}\label{eq:f}
 \boldsymbol{f}_i \equiv \sigpi_i - ( \sigi_i + \varepsilon \nu_i).
\end{equation}
To bound $\boldsymbol{f}_i$, note that 
 $\boldsymbol{f}_i = \boldsymbol{d}_i - \frac{\varepsilon^2}{2} \| \nui_i \|^2 \sigi_i$,
 and therefore for positive integer $k$,
\begin{equation}\begin{aligned}\label{eq:bound_f}
 \sum_{i=1}^N & \E[]{\| \boldsymbol{f}_i \|^{2k}} \\
 & \le \sum_{i=1}^N 4^k \! \left[ \left(\tfrac{\varepsilon^2}{2}\right)^{2k} \E[]{ \| \nui_i \|^{4k}} + \E[]{ \| \boldsymbol{d}_i\|^{2k}} \right] \\
 &\le c(k) ([\Tr(C_N)]^{2k} \varepsilon^{4k} +\Tr(C_N)]^{3k} \varepsilon^{6k}) \\
  &\hspace{1cm} \le c(k) [\Tr(C_N)]^{2k} \varepsilon^{4k}.
\end{aligned}\end{equation}

With the bound \eqref{eq:bound_f}, we proceed to bound $g$ in
\eqref{eq:g}  by bounding its three summations,
\begin{dmath*}
 \E[]{ | g |^4} \le c \left( \E[]{ |\sum_{i=1}^N \frac{\partial H}{\partial \sigi_i} \cdot \boldsymbol{f}_i |^4}  \\ 
+ \E[]{ | \frac N 2 \sum_{i=1}^N (\sigpi_i - \sigi_i) \cdot (\sigpi_{i+1} - \sigi_{i+1} +\sigpi_{i-1}- \sigi_{i-1} )|^4}  \\
 + \E[]{ | N \sum_{i=1}^N (\sigpi_i - \sigi_i) \cdot (\sigpi_i - \sigi_i) |^4} \right) \\
 \le c(k,N) N^8 \sum_i^N \left( \E[]{ \| \boldsymbol{f}_i\|^4} + \E[]{ \| \sigpi_i - \sigi_i \|^8} \right)   . 
 \end{dmath*}
We then conclude that
\begin{dmath} \label{eq:bound_g}
 \E[]{| g |^4} \le c N^9 \varepsilon^8 ( 2 [\Tr(C_N)]^4 + \varepsilon^8 [\Tr(C_N)]^8  ).
\end{dmath}
Together with the fact that $1 \wedge x$ is 1-Lipschitz, we can now bound the remainder $\vec r_2$ appearing in \eqref{R2b} as
\begin{align}
 \E[]{\|\vec r_2\|^2} & \le \E[]{ \| \varepsilon P C_N^{1/2} \w g \|^2}  \notag \\
& \le \varepsilon^2 ( \E[]{ \| P C_N^{1/2} \w \|^4} )^{1/2}  ( \E[]{ \| g \|^4} )^{1/2}  \label{R2}  \\
& \le c \varepsilon^2 \Tr(\bar{C}_N) N^{9/2} \varepsilon^4  [\Tr(\bar{C}_N)]^2. \notag
\end{align}

We now proceed to compute the leading order term of 
$$\begin{aligned}
 \E{ \varepsilon \vec \nu (1 \wedge e^{-\beta \delta H })}  = \varepsilon P C_N^{1/2} \E{\w(1 \wedge e^{-\beta \delta H })}.
\end{aligned}$$
Specifically, we derive the approximation \eqref{w_lemma2} in the main text and bound the error of each approximation used.  The calculation utilizes Lemma 2.4 in \cite{MPS} as stated in Lemma \ref{lm:mps}.

For each component of the noise $w_{i,q}$ with $i=1\dots N$ and $q\in\{x,y,z\}$, we compute the expectation in two steps,
\begin{align*}
 &\E{ w_{i,q}  (1 \wedge e^{-\beta \delta H}) }  \\
 & \hspace{1cm} = \EE\left[ \EE\left[ w_{i,q} (1 \wedge e^{-\beta \delta H}) | \w \backslash w_{i,q} \right]\right],
 \end{align*}
 first taking the expectation over $w_{i,q}$ using the above Lemma, then over the remainder of the components of $\w$.
 To apply Lemma \ref{lm:mps} for the first expectation, take 
\begin{equation}\begin{aligned}
\label{aiqbiq}
 a & = a_{i,q} = - \beta \varepsilon ( (\nabla H)^T P C_N^{1/2})_{i,q} ,\\
 b & = b_{i,q} = -\beta \varepsilon (P C_N^{1/2} \w)^T \nabla H - a_{i,q} w_{i,q}, 
 \end{aligned}\end{equation}
 leaving the need to calculate the error denoted 
 \begin{equation}
 \vec r_3 =  \varepsilon  PC_N^{\frac12} \vec{  \tilde r}_3
 \end{equation}
 with
\begin{equation}\label{Ephi}
(\tilde r_{3})_{i,q} = a_{i,q} \EE\left[ e^{\frac{a_{i,q}^2}{2} + b_{i,q}} \Phi \left( - \frac{b_{i,q}}{|a_{i,q}|} - |a_{i,q}|  \right) \right].
 \end{equation}
 We approximate 
 \begin{equation}\label{Ephiapprox}
 e^{\frac{a_{i,q}^2}{2} + b_{i,q}} \approx 1
 \end{equation}
 and bound the error of this approximation,
\begin{dmath}\label{r3i}
(\tilde{r}_{3,1})_{i,q} \equiv  \ \E{\left( e^{\frac{a_{i,q}^2}{2} + b_{i,q}}-1 \right) \Phi \left( - \frac{b_{i,q}}{|a_{i,q}|} - |a_{i,q}| \right) },   
\end{dmath}
 next.

To bound $(\tilde{r}_{3,1})_{i,q}$, we use that for $ z \sim \N(\mu,\varsigma^2)$, 
\begin{dmath} \label{eq:ezk}
\E[]{\left| e^z -1 \right|^k} = \E[]{\left| e^z -1 \right|^k \mathbbm{1}(z \le 2) }+
\E[]{\left| e^z -1 \right|^k \mathbbm{1}(z > 2)},
\end{dmath}
where the indicator function $\mathbbm{1}$ return one if the statement is true and zero otherwise.
To bound the first term in \eqref{eq:ezk}, since $ |e^z -1| \le e^2 z$ for $z \le 2$, we have that
\begin{dmath*} 
\E[]{\left| e^z -1 \right|^k \mathbbm{1}(z \le 2)| } \le e^{2k} \E[]{ z^k}. 
\end{dmath*} 
To bound the second term in \eqref{eq:ezk}, we note that $\left| e^z -1 \right|^k \le e^{kz}$ when $z >2$, and if $2+\mu+k \varsigma^2 \ge 1$ then we have that
\begin{align*}
&\E[]{e^{kz} \mathbbm{1}(z>2)} 
= \int_2^\infty \frac{1}{\sqrt{2\pi \varsigma^2}} e^{kz} e^{ -\frac{(z-\mu)^2}{2\varsigma^2} } \df z \\
&\hspace{.5cm} = e^{k\mu + \frac{k^2 \varsigma^2}{2}} \int_{2+\mu+k \varsigma^2}^{\infty} \frac{1}{\sqrt{2\pi \varsigma^2}}  e^{-\frac{x^2}{2\varsigma^2}} \df x \\
\end{align*}
under the change of variables $x = z - \mu - k\varsigma^2$.  This Gaussian integral is bounded by the exponential integral as
$$
\int_{2+\mu+k \varsigma^2}^{\infty} e^{-\frac{x^2}{2\varsigma^2}} \df x \le  \int_{2+\mu+k \varsigma^2}^{\infty}e^{-\frac{x}{2\varsigma^2}} \df x
= 2 \varsigma^2 e^{-\frac{2 + \mu + k \varsigma^2}{2\varsigma^2}} 
$$
and $e^{-(2 + \mu + k \varsigma^2)/2\varsigma^2} \le e^{-1/2\varsigma^2} $.  Therefore, we arrive at the bound
\begin{align*}
\E[]{e^{kz} \mathbbm{1}(z>2)} \le  e^{k\mu + \frac{k^2 \varsigma^2}{2}} \sqrt{\frac{2}{\pi}} \varsigma e^{-\frac{1}{2\varsigma^2}}.
\end{align*}

Notice that $\frac{a_{i,q}^2}{2} + b_{i,q} \sim \N( g_1 \varepsilon^2, g_2 \varepsilon^2)$, where 
\[\begin{aligned}
g_1 & =  \frac 12 \beta^2 ( (\nabla H)^T P C_N^{1/2})_{i,q}^2\\
 g_2 & = (\nabla H)^T P C_N P^T \nabla H - a_{i,q}^2.
\end{aligned}\]
Therefore the condition $2+\mu+k \varsigma^2 = 2 + g_1\varepsilon^2 + k g_2 \varepsilon^2 \ge 1$ is met when we take $\varepsilon$ small enough, and applying the above derived bounds for the two terms in \eqref{eq:ezk}, noting that for $k=2$, $\E[]{ z^k}= \mu^2 + \varsigma^2$,  we arrive at
\begin{dmath*}
\E[]{ \left( e^{ \frac{a_{i,q}^2}{2} + b_{i,q}} -1 \right)^2 } \le
e^8 g_2 \varepsilon^2 + \sqrt{\frac{2 g_2}{\pi}} \varepsilon e^{-\frac{1}{2g_2 \varepsilon^2}} \le c \varepsilon^2(\nabla H)^T P C_N P^T \nabla H  
\end{dmath*}
as the term $e^{-\frac{1}{2g_2 \varepsilon^2}}$ decays faster than any polynomial of $\varepsilon$ as $\varepsilon \to 0$ .  The bound
\begin{equation}\label{err:r3}
 | ({\tilde r}_{3,1})_{i,q} |^2  \le c  \varepsilon^2 (\nabla H)^T P C_N P^T \nabla H 
\end{equation}
follows.

We return to bounding \eqref{Ephi} using approximation \eqref{Ephiapprox} and consider
\begin{dmath*}
  \E{\Phi \left( - \frac{b_{i,q}}{|a_{i,q}|} - |a_{i,q}|  \right)}.
\end{dmath*}
 Since both $a_{i,q}$ and $b_{i,q}$ are both proportional to $\varepsilon$, the ratio $b_{i,q}/|a_{i,q}|$ is large relative to $|a_{i,q}|$ and we approximate
\begin{dmath*}
  \E{\Phi \left( - \frac{b_{i,q}}{|a_{i,q}|} - |a_{i,q}|  \right)} \approx
   \E{\Phi \left( - \frac{b_{i,q}}{|a_{i,q}|} \right)}. 
\end{dmath*}
We bound the error of this approximation,
$$
(\tilde r_{3,2})_{i,q} =  \E{\Phi \left( - \frac{b_{i,q}}{|a_{i,q}|} - |a_{i,q}|  \right)} -  \E{\Phi \left( - \frac{b_{i,q}}{|a_{i,q}|} \right)}
$$
by noting that
\begin{dmath*}
 \left| \Phi \left( - \frac{b_{i,q}}{|a_{i,q}|} - |a_{i,q}|  \right) - \Phi \left( - \frac{b_{i,q}}{|a_{i,q}|} \right) \right| \le  \frac{1}{\sqrt{2\pi}} |a_{i,q}|, 
\end{dmath*}
therefore
\begin{equation}\label{err:r4}
| ({\tilde r}_{3,2})_{i,q} |^2 \le \; c \beta^2 \varepsilon^2 .
\end{equation}

We calculate 
$$
 \E{\Phi \left( - \frac{b_{i,q}}{|a_{i,q}|} \right)} = \frac12
$$
by noting that for  $z \sim N(0,\varsigma^2)$,
\[
    \E[]{\Phi(z)} = \E[]{\left(\Phi(z)-\frac 12 \right) + \frac 12} = \frac 12.  
\]
Retracing our steps, we see that
\[
 \E[]{ w_{i,q} (1 \wedge e^{-\beta \big( H(\sigp{}) - H(\sig) \big) })}  \approx \frac{a_{i,q}}{2}
\]
and  \eqref{w_lemma2} follows.

Thus, $\vec r_3$, the error of the approximation in \eqref{w_lemma2}, is bounded as
\begin{align*}
 & \E{\varepsilon P C_N^{1/2} w (1 \wedge e^{-\beta \varepsilon (PC_N^{1/2} w)^T \nabla H} ) }  \\
 & \hspace{1.75cm} -  \left( -\varepsilon^2 \frac \beta 2 P C_N P^T \nabla H \right),
\end{align*}
then
\begin{align}
 \E{ \| \vec r_3 \|^2} \le c N \varepsilon^6 \beta^6 \Tr(C_N) (\| \nabla H\|^4  + \| \nabla H \|^6)  \label{R4}
\end{align}
so that its components involve $a_{i,q}$, a term of size $\varepsilon$, times the error accumulated in the approximations bounded by $(\tilde r_{3,1})_{i,q}$ in \eqref{err:r3} and $(\tilde r_{3,2})_{i,q}$ in \eqref{err:r4}.

\subsubsection{The It\^o Correction Term \label{append:ito_correction}}

Here we consider the approximation
\begin{align*}
&  \E{ \frac 12 \varepsilon^2 \|\nui_i\|^2 \sigi_i  (1 \wedge e^{-\beta \big( H(\sigp{}) - H(\sig) \big) })}   \\
& \hspace{1cm} \approx  \E{ \frac 12 \varepsilon^2 \|\nui_i\|^2 \sigi_i }
\end{align*}
of the second term on the right-hand side of Eq.~\ref{A1},  bounding the error term given in \eqref{err_r5i}.  This includes computing the right-hand side of \eqref{ito} and showing that it corresponds to the It\^o correction of the Stratonovich SDE \eqref{eq:colorSDE}.

First we approximate $H(\sigp{} ) - H(\sig)$ by $\delta H$ given in \eqref{deltaH} and then compute
\begin{equation}
\label{MHIto}
 \E{ - \frac 12 \varepsilon^2 \|\nui_i\|^2 \sigi_i (1 \wedge e^{-\beta\delta H }\big) }  \approx \E{- \frac 12 \varepsilon^2 \|\nui_i\|^2 \sigi_i} ,
\end{equation}
finding it has the same value regardless if the cross-product projection or the cross-cross-product projection is used to obtain $\nui_i$.
With $u_{i,q}$ defined in \eqref{eq:u}, consider the cross-product projection,   $\nui_i =  \sigi_i \times \boldsymbol{u}_i $,
then
$$\begin{aligned}
& \E{- \frac 12 \varepsilon^2 \|\nui_i\|^2 \sigi_i} \\
  =& - \frac 12 \varepsilon^2 \mathbb{E}_n \Big[ (\sigiq_{i,z} u_{i,y})^2 + (\sigiq_{i,y} u_{i,z})^2 +(\sigiq_{i,x} u_{i,z})^2 \\
  &+ (\sigiq_{i,z} u_{i,x})^2  +(\sigiq_{i,y} u_{i,x})^2 + (\sigiq_{i,x} u_{i,y})^2 \Big] \sigi_i,
  \end{aligned}$$
  where we have used that $u_{i,x}, u_{i,y}$ and $u_{i,z}$ are independent and mean zero.  For the expectation of one $(u_{i,q})^2$ we have that
$$\begin{aligned}
\mathbb{E}_n \Big[  (u_{i,q})^2 \Big] &= \sum_{j=1}^N \mathbb{E}_n\Big[ \lambda_j^2 \phi_{ji}^2 (\wiq_{j,q})^2\Big] = \sum_{j=1}^N \lambda_j^2 \phi_{ji}^2,
\end{aligned}$$
where we have used that the $\wiq_{j,x}, \wiq_{j,y}$ and $\wiq_{j,z}$ are independent and mean zero for each $j=1\dots N$.  Together,
\begin{equation}\label{B17}
\E{- \frac 12 \varepsilon^2 \|\nui_i\|^2 \sigi_i} = - \varepsilon^2 \sum_{j=1}^N \lambda_j^2 \phi_{ji}^2 \sigi_i,
\end{equation}
where we have used the identity $\sigiq_{i,x}^2 + \sigiq_{i,y}^2 + \sigiq_{i,z}^2 = 1$.

Now consider the cross-cross-product projection $\nui_i = - \sigi_i \times ( \sigi_i \times \boldsymbol{u}_i) $.  Note that 
$$
\|\nui_i\|^2=\|  - \sigi_i \times ( \sigi_i \times \boldsymbol{u}_i) \|^2 = \|  \sigi_i \times \boldsymbol{u}_i \|^2
$$ 
under the assumption that $\| \sigi_i \|^2=1$.  Therefore, for $\nui_i = - \sigi_i \times ( \sigi_i \times \boldsymbol{u}_i) $
Eq.~\eqref{B17} above also holds.

Returning to \eqref{err_r5i}, we bound the error by
\begin{dmath*}
 \E[]{ \| (\boldsymbol{r_{4}})_i \|^2 } \\
 \le ( \E[] { (- \frac 12 \varepsilon^2 \|\nui_i\|^2 \sigi_i)^4  })^{1/2}  (\E[]{ (-\beta \delta H)^4  })^{1/2} \\
 \le c \varepsilon^6 ( \E[]{ \| \nui_i \|^8 })^{1/2} [ (\nabla H)^T P C_N P \nabla H]^2 .
\end{dmath*}
Bounding the magnitude of this vector, we have that
\begin{dmath}
\label{R5}
 \E[]{ \| \vec r_4 \|^2 } \le c \varepsilon^6 \Tr(\bar{C}_N) [ (\nabla H)^T P C_N P \nabla H ] ^2 
\end{dmath}
regardless of the block projection matrix used, $P$.

Last, we will calculate the It\^o correction for \eqref{eq:colorSDE} and show it is equivalent to the right-hand side of \eqref{ito}.  For the Stratonovich SDE of the form
$d X_t = \mu(t,X_t) dt + B(t,X_t) \circ dW_t$, the corresponding It\^o SDE is~\cite{oksendal2003stochastic}
\[
 d X_t = \tilde{\mu}(t,X_t) dt + B(t,X_t) dW_t,
\]
where 
\[
 \tilde{\mu}_i (t,x) = \mu_i(t,x) + \frac 12 \sum_j \sum_k \frac{\partial B_{ij}}{\partial x_k} B_{kj}.
\]
Ignoring the  constant coefficient $\sqrt{2/\beta}$ for now, the Stratonovich SDE for a single spin taken from \eqref{eq:colorSDE} can be written as
\begin{equation} \label{eq:sde_xyz}
\begin{aligned}
 d \sigiq_{i,x} &= \mu_{i,x} dt + \sigiq_{i,y} d U_{i,z} -  \sigiq_{i,z} d U_{iz} \\
 d \sigiq_{i,z} &= \mu_{i,z} dt + \sigiq_{i,x} d U_{i,y} - \sigiq_{i,y} d U_{ix}
\end{aligned}
\end{equation}
where $d U_{i,q} = \sum_{j=1}^N \lambda_j \phi_{ji} d W_{j,q}$ for $q\in\{x,y,z\}$.  Consider first $\sigiq_{i,x}$ 
\[\begin{array}{lll}
 B_{ix,jx} = 0, & B_{ix,jy} = - \sigiq_{i,z} \lambda_j \phi_{ji}, & B_{ix,jz} = \sigiq_{i,y} \lambda_j \phi_{ji} \end{array}.
\]
Since $B_{ix,jx}=0$, all the partial derivatives in the It\^o correction are zero. For $B_{ix,jy}$, only $\frac{\partial B_{ix,jy}}{\partial \sigiq_{i,z}} = - \lambda_j \phi_{ji} \ne 0$ and the corresponding $B_{iz,jy}$ are $ \sigiq_{i,x} \lambda_j \phi_{ji}$. Therefore, we have that 
\begin{dmath*}
 \sum_{j=1}^N \frac{\partial B_{ix,jy}}{\partial \sigiq_i^z} B_{iz,jy}  = - \sum_{j=1}^N \lambda_j^2 \phi_{ji}^2 \sigiq_{i,x}.
\end{dmath*}
Similarly, for $B_{ix,jz}$, only $\frac{\partial B_{ix,jz}}{\partial \sigiq_{i,y}} = \lambda_j \phi_{ji} \ne 0$ and the corresponding $B_{iy,jz} = -\sigiq_{i,x} \lambda_j \phi_{ji}$.  Therefore, we have that
\begin{dmath*}
 \sum_{j=1}^N \frac{\partial B_{ix,jz}}{\partial \sigiq_{i,y} } B_{iy,jz}  = - \sum_{j=1}^N \lambda_j^2 \phi_{ji}^2 \sigiq_{i,x}.
\end{dmath*}

Above is for cross product.
For cross cross product,
\begin{align*}
  & B_{ix,jx} = (1-\sigma_{ix}^2) \lambda_j \phi_{ji} \\
  & B_{ix,jy} = - \sigma_{ix} \sigma_{iy} \lambda_j \phi_{ji}\\ & B_{ix,jz} = - \sigma_{ix,iz} \lambda_j \phi_{ji}
\end{align*}
and $B_{iy,jq},B_{iz,jq}$ follow similarly.
For $\sigma_{ix}$,
\begin{align*}
 \frac{\partial B_{ix,jx}}{\partial \sigma_{ix}} B_{ix,jx} &= - 2 \sigma_{ix} (1-\sigma_{ix}^2) \lambda_j^2 \phi_{ji}^2 \\
 \frac{\partial B_{ix,jy}}{\partial \sigma_{ix}} B_{ix,jy} &= - \sigma_{iy} (- \sigma_{ix}\sigma_{iy}) \lambda_j^2 \phi_{ji}^2 \\ 
 \frac{\partial B_{ix,jy}}{\partial \sigma_{iy}} B_{iy,jy} &= - \sigma_{ix} (1 -\sigma_{iy}^2) \lambda_j^2 \phi_{ji}^2 \\ 
 \frac{\partial B_{ix,jz}}{\partial \sigma_{ix}} B_{ix,jz} &= - \sigma_{iz} (- \sigma_{ix}\sigma_{iz}) \lambda_j^2 \phi_{ji}^2 \\ 
 \frac{\partial B_{ix,jz}}{\partial \sigma_{iz}} B_{iz,jz} &= - \sigma_{ix} (1 -\sigma_{iz}^2) \lambda_j^2 \phi_{ji}^2 
\end{align*}
Summing the above, we have for $\sigma_{ix}$ the It\^o correction is
\begin{align*}
& \frac 12 \sum_j \lambda_j^2 \phi_{ji}^2 (- \sigma_{ix}) \\
 &\times \Big[ 2 (1-\sigma_{ix}^2) + (1- \sigma_{iy}^2) + (1- \sigma_{iz}^2) - \sigma_{iy}^2 - \sigma_{iz}^2 \Big]   \\
 & = - \sum_j \lambda_j^2 \phi_{ji}^2 \sigma_{ix}
\end{align*}

The above shows that the \Ito form of the SDE has drift coefficient
\begin{equation}\label{ito_correction}
 \tilde{\mu}_{ix} = \mu_{ix} - \sum_{j=1}^N \lambda_j^2 \phi_{ji}^2 \sigiq_{i,x}.
\end{equation}
The calculations for $\tilde{\mu}_{i,y}$ and $\tilde{\mu}_{i,z}$ follow similarly.  

With $ \sum_{j=1}^N \lambda_j^2 \phi_{ji}^2 = \frac{1}{N}\Tr(\bar{C}_N)$ for each $i$, and adding in the coefficient $\sqrt{2/\beta}$, the \Ito drift for the equivalent equation to \eqref{eq:colorSDE} is 
\begin{equation}
 \tilde{\mu} = \mu - \frac{2}{\beta} \frac{1}{N} \Tr(\bar{C}_N) \sig.
\end{equation}
Recall the time-change to arrive at the SDE, $ \delta t = \beta \epsilon^2 / 2 $, with which we see that the above addition to the drift is equivalent to the calculated term in \eqref{B17}.


\subsection{The Diffusion of One Metropolis Hastings Step\label{App:Diff}}

In this section, we bound the error of approximating the diffusion part of one step of the MH algorithm as in Eq.~\ref{eq:diff_approx} in the main text,
\begin{equation}\label{phierrordef}
\phib_i^n \equiv \sigi_i^{n+1} - \sigi_i^n - \E{\sigi_i^{n+1} - \sigi_i^n} - \varepsilon \nui_i^n.
\end{equation}
 This random variable $\phib_i^n$ takes the values
\begin{dmath*}
 \phib_i^n = 
 \begin{cases}
  \boldsymbol{ f}_i - \E{\sigi_i^{n+1} - \sigi_i^n} & \textrm{with prob. } \bar{\alpha} \\
  - \E{\sigi_i^{n+1} - \sigi_i^n} - \varepsilon \nui_i^n & \textrm{with prob. } 1 - \bar{\alpha}
 \end{cases}
\end{dmath*}
where $\bar{\alpha} = 1 \wedge e^{-\beta \delta H}$, $\boldsymbol{f}_i$ is defined in Eq.~\ref{eq:f} and $\delta H$ is given in Eq.~\ref{deltaH}.  This error can be bounded by 
\begin{dmath}
\label{Zetan}
 \E[]{ \| \phib_i^n \|^2 } = \E[]{\| \boldsymbol{f}_i^n - \E{\sigi_i^{n+1} - \sigi_i^n}\| (1 \wedge e^{-\beta \delta H}) } \\
 + \E[]{ \| \E{\sigi_i^{n+1} - \sigi_i^n} - \varepsilon \nui_i^n \|^2 ( 1 - ( 1 \wedge e^{-\beta \delta H}))) } \\
 \le \E[]{\| \boldsymbol{f}_i^n\|^2} + \left(\E[]{ (\E{\sigi_i^{n+1} - \sigi_i^n} -\varepsilon \nui_i^n )^4}\right)^{1/2} \notag  \\
 \left( \E[]{{ (-\beta \delta H)^2 }} \right)^{1/2}
 \le c \varepsilon^3.  
\end{dmath}
The covariance of the error at different time steps $n>m$ when is
\begin{align} \label{eq:cond_independence}
 \E[]{\zeta_{i,p}^n \zeta_{i,q}^m} 
& = \E[]{\E{\zeta_{i,p}^n \zeta_{i,q}^m}} = \E[]{ \zeta_{i,p}^m \E{\zeta_{i,q}^n}} \nonumber \\
& = \E[]{ \zeta_{i,p}^m \cdot 0 } = 0
\end{align}
for any $i=1\dots N$ and $p,q\in\{x,y,z\}$.


\subsection{Completion of the Proof \label{App:Existence}}

 The \Ito SDE \eqref{eq:colorSDE_ito} has a unique solution before proceeding in the next section to bound the error between the Metropolis Hastings dynamics and this unique SDE solution.  We apply Theorem 5.2.1 in \cite{oksendal2003stochastic} for an (It\^o) SDE of the form $dx = \mu(x)dt + B(x)dW$ by showing the SDE coefficients
\begin{equation}\label{muB}
\begin{aligned}
 \mu(x) &= \frac 1N P_x C_N P_x^T \Delta_N x -  2 \beta^{-1}  \frac{1}{N}  \Tr(\bar{C}_N) x, \\
 B(x) &= \sqrt{\frac{2}{\beta}} P_x C_N^{1/2}.
 \end{aligned}
\end{equation}
are Lipschitz continuous, which is a relatively straightforward calculation.  There is an analogous argument in \cite{gao2018limiting}, Section $3$.   

Following further the convergence results in \cite{gao2018limiting}, Section $3$, we can complete the proof of Theorem \ref{thm:MHtoSDE} \label{App:proof}.  The proof is similar to the proof of the Stochastic Euler method. We will first prove a bound for the strong error
\begin{equation}
 \tilde{e}(t) = \E[]{\| \s(t) - \sig(t) \|^2}
\end{equation}
at a fixed time $t$, where $\sig(t)$ is the piecewise constant interpolation of the MH dynamics and $\s(t)$ is the solution to the SDE \eqref{eq:colorSDE_ito}.  Then $\tilde{e}(t)$ and Doob's martingale inequality are used to obtain a uniform bound on
\begin{equation}
 e(t) = \E[]{ \sup_{0 \le \tau \le t} \| \s(\tau) -\sig(\tau) \|^2}.
\end{equation}

One must use the It\^o isometry and H\"older's inequality to prove the following Gr\"onwall inequality
\begin{equation} \label{ineq:tilde_e}
 \tilde{e}(t) \le (c_1 t + c_2) \int_0^t e(\tau) \df \tau + c_3 \sqrt{\delta t},
\end{equation}
where $c_1,c_2,c_3$ are functions of $J,N,\beta,\Tr(C_N)$ and this gives the bound
\begin{dmath}
 \tilde{e}(t) \le c_3 \sqrt{\delta t} e^{c_1 t + c_2}.
\end{dmath}

For a fixed $t$, take $ n = \lfloor \frac{t}{\delta t} \rfloor$, then
\begin{equation}\begin{aligned} \label{eq:s_sigma}
 \s(t) &- \sig(t) = \int_0^{n \delta t} \Big[ \mu(\s(\tau)) - \mu(\sig(\tau))\Big]  \df \tau \\
 & +  \int_0^{n \delta t}\Big[  B(\s(\tau)) - B(\sig(\tau)) \Big] \df W_\tau +  \int_{n \delta t}^t \mu(\s(\tau)) \df \tau \\
 & +  \int_{n \delta t}^t B(\s(\tau)) \df W_\tau + \sum_{k=1}^n \vec r^{\;k} + \sum_{k=1}^n \vec\zeta^{\;k},
\end{aligned}\end{equation}
where the drift and diffusion coefficients, $\mu(x)$ and $B(x)$ are given in \eqref{muB} and the errors $\vec r$ and $\vec \zeta$ are bounded in \eqref{Rn}  and \eqref{Zetan}.  The remaining details are almost identical to those in \cite{gao2018limiting}, Section $3.4$ and we refer the reader there for further details.

\section{Invariance of the Gibbs Distribution} \label{append:FKP}

In this appendix, we present some direct calculations showing the invariance of the Gibbs distribution.  In App.~\ref{App:FP_Addative}, we present the well-known case of SDE \eqref{eq:SDEgeneric} with additive noise.  In App.~\ref{App:FP_Mult}, we present the cases of the spin-system SDE \eqref{eq:ItoWhite} with white multiplicative noise using either the $\sigma \times \cdot$ or the $-\sigma \times ( \sigma \times \cdot)$ projection, as well as the case of the SDE \eqref{eq:colorSDE} with colored multiplicative noise, for which only the $\sigma \times$ projection results in the invariance of the Gibbs distribution.

\subsection{Additive Noise\label{App:FP_Addative}}

For the $N$-dimensional SDE \eqref{eq:SDEgeneric} with constant matrix $B$, we show the invariance of the Gibbs distribution $\rho(x) = e^{-\beta H(x)}$ by direct substitution into the Fokker-Planck Equation \eqref{eq:constB_FP}.  As $\partial_j \rho = -\beta (\partial_j H) \rho$, we have that 
\begin{dmath*}
0 
=  \sum_i \sum_k\sum_j B_{ik}B_{jk}\left[ (\partial_i \partial_j H ) \rho -\beta (\partial_j H ) (\partial_i H)\rho \right]
+ \beta^{-1} \sum_i \sum_j   \sum_k B_{ik} B_{jk}\partial_i \left( \partial_j(-\beta H) \rho \right) 
=  \sum_i \sum_k\sum_j B_{ik}B_{jk}\left[ (\partial_i \partial_j H ) \rho -\beta (\partial_j H ) (\partial_i H)\rho \right]
- \sum_i \sum_j   \sum_k B_{ik} B_{jk} \left[ ( \partial_i\partial_j H) \rho  -\beta   (\partial_j H) (\partial_i H)\rho \right]
\end{dmath*}
and the terms on the right-hand side clearly cancel.

\subsection{Multiplicative noise\label{App:FP_Mult}}

Consider the following Stratonovich SDE with multiplicative noise
\begin{equation}\label{StratSDE}
dX = - B(X) B^T(X) \nabla H(X) + \sqrt{2\beta^{-1}} B(X) \circ d W
\end{equation}
where $B$ could be for example the block projection matrices $P_1$ or $P_2$.  It could also be the combination of $P C^{1/2}$.
For the Fokker-Planck equation
\begin{dmath}\label{eq:mult_FP}
\partial_t \rho(x,t) =  \sum_i \partial_i\left\{ \left( B B^T \nabla H \right)_i \rho(x,t)\right\} 
+ \beta^{-1} \sum_k \sum_i \partial_i \left\{ B_{ik} \sum_j \partial_j \left( B_{jk} \rho(x,t)\right) \right\} 
 =  \sum_i \partial_i\left\{  \sum_k\sum_j B_{ik} B_{jk}(\nabla H)_j   \rho(x,t)\right\} 
+ \beta^{-1} \sum_k \sum_i \partial_i \left\{ B_{ik} \sum_j \partial_j \left( B_{jk} \rho(x,t)\right) \right\} ,
\end{dmath}
we consider the invariance of the Gibbs distribution $\rho(x) = e^{-\beta H(x)}$ by direct substitution.
The third line (drift terms) of the above equation leads to the terms
\begin{equation}\begin{aligned}\label{drift_terms}
 \partial_i &\left( B_{ik} B_{jk}\right) (\partial_j H) \rho 
 +  B_{ik} B_{jk}(\partial_i\partial_j H) \rho \\
&  + B_{ik}B_{jk} (\partial_j H) (\partial_i\rho )\\
= &  \partial_i \left( B_{ik} B_{jk}\right) (\partial_j H) \rho 
 +  B_{ik} B_{jk}(\partial_i\partial_j H) \rho \\
&  -\beta B_{ik}B_{jk} (\partial_j H) (\partial_i H) \rho 
\end{aligned}\end{equation}
that are summed over $i,j$ and $k$.
The fourth line (diffusion terms) of Eq.~\ref{eq:mult_FP} leads to the terms
\begin{dmath}\label{diff_terms}
\partial_i \left[ B_{ik} (\partial_j B_{jk}) \rho \right] + 
\partial_i\left[ B_{ik} B_{jk}\partial_j \rho\right] 
= (\partial_i B_{ik})( \partial_j B_{jk} )\rho + B_{ik}(\partial_i \partial_j B_{jk} )\rho +  B_{ik} (\partial_j B_{jk} ) (\partial_i \rho) 
+ \partial_i \left( B_{ik}B_{jk}\right) (\partial_j \rho) + B_{ik} B_{jk} ( \partial_i \partial_j \rho)
= (\partial_i B_{ik})( \partial_j B_{jk} )\rho + B_{ik}(\partial_i \partial_j B_{jk} )\rho -\beta  B_{ik} (\partial_j B_{jk} ) (\partial_i H) \rho 
-\beta \partial_i \left( B_{ik}B_{jk}\right) (\partial_j H) \rho 
-\beta B_{ik} B_{jk} ( \partial_i \partial_j H) \rho +\beta^2 B_{ik} B_{jk} ( \partial_i H) (\partial_j H) \rho
\end{dmath}
multiplied by $\beta^{-1}$ and summed over $i,j$ and $k$.
Combining the terms in \eqref{drift_terms} and \eqref{diff_terms}, the following terms, summed over $i,j$ and $k$, are left over:
\begin{dmath}\label{leftover}
 \beta^{-1} \left\{ (\partial_i B_{ik})( \partial_j B_{jk} )\rho + B_{ik}(\partial_i \partial_j B_{jk} )\rho\right\}
 -  B_{ik} (\partial_j B_{jk} ) (\partial_i H) \rho .
\end{dmath}
For a generic multiplicative noise in the SDE of the form \eqref{StratSDE}, the Gibbs distribution is not guaranteed to be an invariant measure.  Next, we consider specific cases for the matrix $B$.

We first consider the case of a single spin, $X\in\mathbb{S}^2$, and the matrix $B$ as either the $3\times 3$ projection matrix $P_1$ in  Eq.~\ref{eq:Pcross} or $P_2$ in  Eq.~\ref{eq:Pcrosscross}.  For the case of $P_1$, the terms in \eqref{leftover} are all zero,
$$\begin{aligned}
\sum_{i,j,k} (\partial_i B_{ik})( \partial_j B_{jk} ) = 0 \\
\sum_{i,j,k} B_{ik}(\partial_i \partial_j B_{jk} ) = 0 \\
\sum_{i,j,k}  B_{ik} (\partial_j B_{jk} ) (\partial_i H) = 0
\end{aligned}$$
and we conclude that the Gibbs distribution is invariant.  For the case of $P_2$, the terms in \eqref{leftover} are
$$\begin{aligned}
\sum_{i,j,k} (\partial_i B_{ik})( \partial_j B_{jk} ) = 4(\sigma_x^2 + \sigma_y^2 + \sigma_z^2) &= 4 \\
\sum_{i,j,k} B_{ik}(\partial_i \partial_j B_{jk} ) = -4(\sigma_x^2 + \sigma_y^2 + \sigma_z^2) &= -4 \\
\sum_{i,j,k}  B_{ik} (\partial_j B_{jk} ) (\partial_i H) &= 0.
\end{aligned}$$
The order $\beta^{-1}$ terms cancel while the order one term is zero, thus for the case of $P_2$, the Gibbs distribution is also invariant.  This direct calculation easily extends to the case of $N$ spins with white noise, and we conclude that with either projection in the SDE \eqref{eq:ItoWhite}, the Gibbs distribution \eqref{eq:Gibbs} is invariant.

We now consider the case of the colored noise SDE \eqref{eq:colorSDE}.  Taking $B = P_1 C_N^{1/2}$ and indexing the vector $X$ as $X_{ix} = x_i$, $X_{iy}=y_i$ and $X_{iz} = z_i$, for $i=1\dots N$ with the $i$th spin vector being denoted as $\left<x_i,y_i,z_i\right>$,  we have that
$$
\begin{array}{lll}
 B_{ix,jx} = 0 & B_{ix,jy} = -z_i \lambda_j \phi_{ji} & B_{ix,jz} = y_i \lambda_j \phi_{ji} \\
  B_{iy,jx} = -z_i \lambda_j \phi_{ji} & B_{iy,jy} = 0 & B_{iy,jz} = x_i \lambda_j \phi_{ji} \\
  B_{iz,jx} = y_i \lambda_j \phi_{ji} & B_{iz,jy} = -x_i \lambda_j \phi_{ji} & B_{iz,jz} = 0
  \end{array}
$$
where $\sum_{k=1}^N \bar{C}_{ik}\phi_{kj} = \lambda_j^2 \phi_{ij}$ for $i,j=1\dots N$.  By inspection we see that
$$
\partial_{iq} B_{iq, jp} = 0
$$
for all $p,q\in{x,y,z}$ and therefore all the terms in \eqref{leftover} are zero.  For the case of colored noise and the $\sigma \times$ projection matrix $P_1$, the Gibbs distribution is an invariant measure for the SDE \eqref{eq:colorSDE}.

Taking $B = P_2 C_N^{1/2}$ with $P_2$ given by \eqref{eq:Pcrosscross}, 
\begin{align*}
& B_{ix,jx} = (1-x_i^2) \lambda_j \phi_{ji}, \   B_{ix,jy} = -x_i y_i \lambda_j \phi_{ji} , \\
 & B_{ix,jz} = -x_iz_i \lambda_j \phi_{ji}, \  B_{iy,jx} = -x_iy_i \lambda_j \phi_{ji}  , \\
 &  B_{iy,jy} = (1-y_i^2) \lambda_j \phi_{ji}, \   B_{iy,jz} = -y_iz_i\lambda_j \phi_{ji}  , \\
 & B_{iz,jx} = -x_iz_i \lambda_j \phi_{ji}, \  B_{iz,jy} = -y_iz_i \lambda_j \phi_{ji} ,  \\
& \hspace{1cm}    B_{iz,jz}  = (1-z_i^2) \lambda_j \phi_{ji}  
  \end{align*}
and noting that 
$$\begin{aligned}
\partial_{iq} B_{iq, jq} &= -2 q_i \lambda_j \phi_{ji} \\
\partial_{iq} B_{iq, jp} &= - p_i \lambda_j \phi_{ji} \\
\partial_{jq} H &= - N (q_{j+1}-2q_j + q_{j-1})
\end{aligned}$$
further algebra leads to the conclusion that the third term in \eqref{leftover},
$$
\sum_{i,j,k=1}^N \sum_{p,q,r \in\{x,y,z\}} B_{ip,kr} (\partial_{jq} B_{jq,kr}) \partial_{ip} H \ne 0.
$$
Since this term can never cancel with the order $\beta^{-1}$ terms for arbitrary $\beta$, we conclude that the Gibbs distribution is not an invariant measure for the colored noise SDE \eqref{eq:colorSDE} with the $- \sigma \times ( \sigma \times \cdot)$ projection matrix $P_2$.

\section{Well-posedness of the Colored Noise SPDE model for trace class multipliers in Sobolev spaces}
\label{A:LWP}

To establish the local well-posedness of \eqref{eqn:spde} result in the case that $M_\kappa (D)$ is trace class ($\kappa$ sufficiently large), we may write the SPDE in the It\^o formulation
 \begin{align}
\label{eqn:spdeito}
d \sigma & = \left[ -\sigma \times (M_\kappa (D)) (\sigma \times \Delta \sigma)  + \int K_\kappa (x,y) dy \sigma(x) \right] dt  \\
& \hspace{1cm} + \sigma \times ( \mathcal{F}^{-1} (m(|k|))^{-\kappa} dW (k). \notag
\end{align}
Note that the lack of any geometric projection in the It\^o correction term follows from a cancellation that arises from direct computation very similar to that in \ref{append:ito_correction}.  To prove local well-posedness of \eqref{eqn:spdeito}, we establish some baseline energy estimates by first mollifying the equation, then proving a priori bounds in a high enough Sobolev space.  

To proceed, let us record a few useful facts.  First,  $H^s (\bbT^d)$ is an algebra for $s > d/2$.  Namely, $\| u v \|_{H^s} \leq \|u \|_{H^s} \| v \|_{H^s}$ for $s$ sufficiently large.  Secondly, $[\partial_x , (I - \Delta)^{\kappa/2}] = 0$.  Lastly, we observe that
\begin{eqnarray}
\label{H1est}
& \int \Delta \sigma  \cdot (-\sigma \times (M_\alpha (D)) (\sigma \times \Delta \sigma) )d x \\
& \hspace{1cm} = \int (\sigma \times \Delta \sigma )  \cdot  (M_\alpha (D)) (\sigma \times \Delta \sigma) dx > 0, \notag
\end{eqnarray}
which allows us to generate a priori bounds on $\| \sigma \|_{H^s}$ uniformly bounded in $\epsilon$ for $s > \frac{d}{2}$.  Note, this also shows that a classical solution will have decaying $H^1$ norm, as would be expected from the structure of the Gibbs measure.  

\begin{remark}
We can easily see that were we using the cross-cross projection on the drift term, then we could write the deterministic flow generated by the drift as
\[
\partial_t \sigma = M_\kappa (D) \Delta \sigma + M_\kappa (D) ( | \nabla \sigma|^2 \sigma)
\]
suggesting that the dynamics of this non-local PDE should have a different diffusion time scale roughly given by $M_\alpha( \lambda_1) \lambda_1^2$.  Numerically we observe a similar diffusion scaling for the cross projection, which is explored in Section \ref{sec:num} in the main text.  
\end{remark}

The outline of the local well-posedness argument proof a la \cite{Tay3}, Ch. 15 proceeds as follows:
 \begin{itemize}
  \item Define $\sigma_\epsilon = \chi(D/\epsilon) \sigma$, a frequency cut-off version of the equation. 
 \item By the \eqref{H1est} adapted to this setting, this ODE system has global existence for each $\epsilon$
 \item For $s > \frac{d}{2}+2$, computing $\partial_t \mathbb{E}\left[ \| \sigma_\epsilon \|_{\dot{H}^s}^2 \right]$ in a similar fashion to \eqref{H1est} gives a signed quantity on the highest derivatives and using the algebra property gives uniform a uniform existence time bound using a simple Gronwall inequality argument in $\epsilon$ on $\mathbb{E}\left[ \| \sigma_\epsilon \|_{H^s} \right]$ provided $s$ is sufficiently large provided the initial data is sufficiently regular.   We use the It\^o formula 
 \begin{align*}
 & \mathbb{E}\left[ \| \sigma_\epsilon \|_{\dot{H}^s}^2 \right]     = \mathbb{E}\left[ \| \sigma_\epsilon \|_{\dot{H}^s}^2 (0) \right] \\
 & + \mathbb{E} \left[ \int_0^t \langle \sigma_\epsilon (r) , \sigma_\epsilon \times (M_\kappa (D)) (\sigma_\epsilon \times \Delta \sigma_\epsilon) (r) \rangle_{\dot{H}^s} dr  \right]\\
 & + \mathbb{E} \left[ \int_0^t \langle \sigma_\epsilon (r) , \int K_\kappa (x,y) dy \sigma_\epsilon (x,r) \rangle_{\dot{H}^s} dr \right] \\
&  +  \mathbb{E} \left[ \int_0^t \| \sigma_\epsilon \times  ( \mathcal{F}^{-1} (m(|k|))^{-\kappa}   \|_{L_2 (L^2,\dot{H}^s)}^2  dr   \right],
\end{align*}
where $L_2( H_1, H_2)$ is the space of linear Hilbert-Schmidt operators from Hilbert space $H_1$ to Hilbert space $H_2$.
 \item Since balls in the $H^s$ norm are compact, the process is tight and hence 
 taking the weak limit in $\epsilon$ gives the solution $\sigma \in H^s$.
 \item This can be bootstrapped into supremum over time.
 \item Uniqueness and continuity follow from energy estimates once the classical solution has been constructed.  
 \item  Note, in all of this, there is an It\^o correction  term to be controlled in each energy estimate.  However, as it carries no derivatives, it is lower order and can be absorbed in the energy estimate without much difficulty. 
 \end{itemize}

\begin{remark}
This rather crude treatment of \eqref{eqn:spde} is by no means the state of the art.  However, as we are more concerned here with the microscopic to macroscopic convergence to such an equation, it suffices for our purposes.  The recent work of \cite{bruned2019geometric} handles the white noise case of \eqref{eqn:spde} and gives an appropriate re-normalization technique adapted to a general geometric setting.  It would be interesting to extend this to the range of covariance matrices given here between identity (white) and $\kappa > \frac12$ (trace class), but this pursuit goes beyond the focus of the present work.  
\end{remark}

\subsection*{Acknowledgements}

Y.G. and J.L.M were supported in part by NSF Applied Math Grant DMS-1312874 and NSF CAREER Grant DMS-1352353.  J.C.M. was partially supported by DMS-1613337 from the National Science Foundation and a Simons Travel grant.  The authors thank Amarjit Budhiraja for helpful conversations during the preparation of this work.   J.L.M. acknowledges Duke University and MSRI where he was hosted during part of the completion of this project.

\bibliography{color,spin}

\end{document}